\documentclass[twocolumn,twocolappendix,tighten,usenatbib,useAMS,fleqn,usenames,dvipsnames]{aastex63}
\usepackage{amsmath}
\usepackage{cancel}

\graphicspath{{./}{figures/}}
\received{\today}
\revised{}
\accepted{}
\submitjournal{ApJL}

\definecolor{internationalkleinblue}{rgb}{0.0, 0.4, 0.8}
\definecolor{britishracinggreen}{rgb}{0.0, 0.5, 0.3}
\hypersetup{colorlinks=true, citecolor=internationalkleinblue, linkcolor=internationalkleinblue, urlcolor=internationalkleinblue}

\newcommand{\eg}{e.g.}

\newcommand{\lperp}{\ell_\perp}
\newcommand{\lpar}{\ell_\parallel}
\newcommand{\vecl}{\pmb{\ell}}

\newcommand{\vA}{v{\mskip-3mu}_{A}}
\newcommand{\vrec}{v{\mskip -1mu}_{\rm r{\mskip -1mu}e{\mskip -1mu}c}}
\newcommand{\vshear}{v{\mskip -2mu}_{\rm s{\mskip -1mu}h{\mskip -1mu}e{\mskip -1mu}a{\mskip -1mu}r}}
\newcommand{\Brec}{B_{\rm rec}}
\newcommand{\Bguide}{B{\mskip -2mu}_{\rm g{\mskip -1mu}u{\mskip -1mu}i{\mskip -1mu}d{\mskip -1mu}e}}

\newcommand{\etaM}{\eta}%_{\rm M}}

\newcommand{\curl}{\boldsymbol{\nabla} \times}

\newcommand{\vecv}{\pmb{v}}
\newcommand{\vecB}{\pmb{B}}
\newcommand{\vecj}{\pmb{j}}

\newcommand{\Mach}{\mathcal{M}}
\newcommand{\MachA}{\mathcal{M}_A}

\shorttitle{ISM Plasmoids}
\shortauthors{Fielding, Ripperda, \& Philippov}

\begin{document}

\title{Plasmoid Instability in the Multiphase Interstellar Medium}

\correspondingauthor{Drummond B. Fielding}
\email{drummondfielding@gmail.com}

\author[0000-0003-3806-8548]{Drummond B. Fielding}
\affiliation{Center for Computational Astrophysics, Flatiron Institute, 162 5th Ave, New York, NY 10010, USA}
\author[0000-0002-7301-3908]{Bart Ripperda}
\affiliation{School of Natural Sciences, Institute for Advanced Study, 1 Einstein Drive, Princeton, NJ 08540, USA}
\affiliation{NASA Hubble Fellowship Program, Einstein Fellow}
\affiliation{Department of Astrophysical Sciences, Peyton Hall, Princeton University, Princeton, NJ 08544, USA}
\affiliation{Center for Computational Astrophysics, Flatiron Institute, 162 5th Ave, New York, NY 10010, USA}
\author[0000-0001-7801-0362]{Alexander A. Philippov}
\affiliation{Department of Physics, University of Maryland, College Park, MD 20742, USA}

\begin{abstract}
The processes controlling the complex clump structure, phase distribution, and magnetic field geometry that develops across a broad range of scales in the turbulent interstellar medium remains unclear. Using unprecedentedly high resolution three-dimensional magnetohydrodynamic simulations of thermally unstable turbulent systems, we show that large current sheets unstable to plasmoid-mediated reconnection form regularly throughout the volume. The plasmoids form in three distinct environments: (i) within cold clumps, (ii) at the asymmetric interface of the cold and warm phases, and (iii) within the warm, volume-filling phase. We then show that the complex magneto-thermal phase structure is characterized by a predominantly highly magnetized cold phase, but that regions of high magnetic curvature, which are the sites of reconnection, span a broad range in temperature. Furthermore, we show that thermal instabilities change the scale dependent anisotropy of the turbulent magnetic field, reducing the increase in eddy elongation at smaller scales. Finally, we show that most of the mass is contained in one contiguous cold structure surrounded by smaller clumps that follow a scale free mass distribution. These clumps tend to be highly elongated and exhibit a size versus internal velocity relation consistent with supersonic turbulence, and a relative clump distance-velocity scaling consistent with subsonic motion. We discuss the striking similarity of \emph{cold plasmoids} to observed tiny scale atomic and ionized structures and H\textsc{i} fibers, and consider how the presence of plasmoids will modify the motion of charged particles thereby impacting cosmic ray transport and thermal conduction in the ISM and other similar systems.
\end{abstract}

\keywords{Interstellar magnetic fields (845), Interstellar medium (847), Interstellar plasma (851), Plasma astrophysics (1261), Galaxy formation (595), Galaxy magnetic fields (604), Cosmic rays (329)}

\section{Introduction} \label{sec:intro} 
The interstellar medium (ISM) is a dynamic and varied environment and is the nexus of numerous phenomena that combine to regulate the formation of stars, the growth of supermassive black holes, and the evolution of galaxies. The ISM is fundamentally inhomogeneous, consisting of multiple components broadly categorized as cold (with temperature $T\lesssim 10^2$ K), warm ($T\sim 10^4$ K), and hot ($T\gtrsim 10^6$ K) plasma that form as a result of thermal instabilities \citep{Field:1969} and supernova (SN) heating \citep{McKeeOstriker:1977}. Adding to the complexity is ubiquitous turbulence and the presence of dynamically important magnetic fields throughout the ISM, which play an essential role in determining the energetics, phase structure, and evolution of these systems \citep[for reviews, see, \eg][]{Elmegreen:2004,Klessen:2016}. Despite the central importance of the ISM to many astrophysical systems a robust theory of the interplay of thermal instabilities, turbulence, and magnetic fields remains elusive.

An understanding of the turbulent magnetized ISM must build upon the extensive work on magnetohydrodynamic (MHD) turbulence \citep[for a comprehensive review, see,][]{Schekochihin:2020}. This is complicated by the multiphase nature of the ISM. Different phases can have very different sonic and Alfv\'enic Mach numbers, which can dramatically change the nature of the MHD turbulence \citep[\eg][]{Burkhart:2009}. It is, therefore, necessary to simultaneously include the effect of strong radiative cooling and heating in this picture. This has been investigated in the purely hydrodynamic case \citep[\eg][]{Vazquez-Semadeni:2000,Seifried:2011,Gazol:2013} and in the presence of magnetic fields \citep{Kritsuk:2017}. These avenues of study have begun to uncover many important features of the ISM phase structure, and the structure of the turbulent magnetic and velocity fields.

Adding to the complexity is the evidence that magnetic reconnection operates in the ISM and that it is essential for the operation of a galactic dynamo \citep{Zweibel:1997}. The ohmic diffusion timescales of typical interstellar magnetic fields are extremely long, which is incompatible with the observed ratio of the magnetic field strength in the small-scale random component to the large-scale ordered component that is ${\sim} 3$ \citep[\eg][]{Heiles:1996}. The problem stems from standard reconnection model predictions that the reconnection rate decreases with resistivity \citep{Parker:1957,Sweet:1958}, and the fact that ISM resistivity is extremely small. A similar need for fast reconnection arises in numerous other astrophysical systems, such as solar and black hole flares \cite[\eg][]{Ripperda:2022, Yan:2022}. 

Recently, it has been shown that fast reconnection is governed by the tearing instability\footnote{See \citep{Lazarian:1999} for an alternative theory in which turbulence accelerates the reconnection rate.}, which results in a reconnection rate that is independent of the resistivity for Lundquist numbers $S = L_{\rm CS} \vA / \etaM$ (where $L_{\rm CS}$ is the length of the current sheet, $\vA$ is the local upstream Alfv\'en velocity, and $\eta$ is the ohmic resistivity) above $S_{\rm crit} {\sim} 10^4$ \citep{Loureiro:2007}. Current sheets with $S \gtrsim 10^4$ will break apart due to the tearing instability \citep{Furth:1963} into smaller current sheet at numerous new ``X-points.'' This locally reduces the Lundquist number to ${\sim}10^4$, and thereby prevents the reconnection rate from dropping below the asymptotic value of $\vrec/\vA = S_{\rm crit}^{-1/2} \sim 0.01$ (where $\vrec$ is the velocity at which reconnecting magnetic fields flow into the current sheet). As a result of this process, also referred to as the plasmoid instability \citep{Bhattacharjee:2009,Uzdensky:2010}, long strings of plasmoids\footnote{In 3D these structures manifest as flux tubes, but we refer to them as plasmoids hereafter.} (secondary islands) are formed. 

Despite the clear need for fast reconnection in the ISM, the plasmoid instability has, to our knowledge, never been studied in realistic multiphase turbulent ISM conditions \citep[see, however,][for magnetic reconnection induced by molecular cloud collisions in idealized isothermal simulations]{Kong:2021,Kong:2022}. In this paper, we endeavor to do just that, and study the formation of tearing unstable current sheets, and their subsequent break up into plasmoids in turbulent thermally unstable multiphase ISM conditions. The Lundquist number in the ISM is expected to exceed the critical value for the onset of the tearing instability by many orders of magnitude. For example, taking $\vA = 10$ km s$^{-1}$ and using Spitzer resistivity $\etaM~\approx~10^{7} (T/10^4 {\rm \, K})^{-3/2}\, {\rm cm}^2 \, {\rm s}^{-1}$ \citep{Spitzer:1953} as is characteristic for the warm ionized medium, all current sheets with lengths greater than $L_{\rm CS} > 1$ km will have $S > S_{\rm crit}$. We can, therefore, confidently expect plasmoid-mediated magnetic reconnection to be a common occurrence the ISM. Solving the visco-resistive MHD equations to simulate this process is, however, challenging because resolving the length scales associated with the plasmoid instability requires extreme resolutions \citep[\eg][]{Dong:2022}.

Here, we present a high resolution\footnote{To our knowledge this is the highest resolution turbulent magnetized simulations of a bistable medium published to date.} three dimensional (3D) driven turbulence MHD simulation that includes ISM-like cooling and heating functions, in service of two main goals. The first goal is to demonstrate the existence of plasmoids in the ISM. To this end, we highlight the various formation pathways demonstrating that, in addition to turbulent folding, thermal instabilities can form plasmoid unstable current sheets. The second goal is to explore the quantitative impact including these various physical processes at very high resolution has on the phase structure, magnetic geometry, and statistical properties. In \autoref{sec:setup} we introduce the numerical simulations, followed by an investigation of their properties on a qualitative then quantitative level in \autoref{sec:results}. In \autoref{sec:discussion} we summarize and discuss the implications of our findings with a particular focus on potential observables, such as extreme scattering events, and on cosmic ray transport. 

Movies of these simulations can be found at \url{https://dfielding14.github.io/movies/}.

%-------------------------------------------%
%-------------------------------------------%
%-------------------------------------------%
%-------------------------------------------%
\section{Numerical Methods} \label{sec:setup}
We use the \texttt{athena++} code framework \citep{athena++} to run a suite of 2.5D and 3D visco-resistive MHD simulations on a static Cartesian mesh with periodic boundary conditions using a second order spatial reconstruction, a third order time integrator, and the HLLD Riemann solver. We adopt an $\mathcal{E} = P / (\gamma - 1)$ equation of state with an adiabatic $\gamma = 5/3$, where $\mathcal{E}$ is the thermal energy and $P$ is the thermal pressure. 

We use a turbulent driving source term in our simulations that follows an Ornstein-Uhlenbeck process with a correlation timescale equal to the largest eddy turnover time. The momentum driving field is chosen such that the Helmholtz decomposition results in a two-to-one ratio between the divergence-free and curl-free components. Turbulence is only driven on large scales, with equal power injected into wavenumbers equal to $k L / 2 \pi = 1$ and $2$. The turbulent kinetic energy injection rate is chosen to yield a unity root-mean-square sonic Mach number, $\Mach^2 = \langle v^2\rangle / \langle c_{\rm s}^2 \rangle =  1$. Note that other somewhat related studies have focused primarily on decaying turbulence (e.g., in 3D \citealt{Chernoglazov:2021,Dong:2022}, and in 2D \citealt{Dong:2018}), as opposed to driven turbulence (see, however, \citealt{Galishnikova:2022}).

The heating rate, $\Gamma$, and cooling curve, $\Lambda$, roughly mimic the heating and cooling in the ISM of the Milky Way \citep[e.g.][]{Koyama:2002}. We sacrifice some degree of physical realism and adopt a simple piecewise power law cooling curve so as to isolate the relevant cooling physics while remaining as general as possible. The heating and cooling curves are designed to ensure that at the initial pressure $P_0$ there is an unstable equilibrium at temperature $T_0$, and two stable equilibria at $T_{\rm cold} = T_0/100$ and $T_{\rm warm} = 10 T_0$.\footnote{For realistic Milky Way like conditions these would take the value of $P_0 \approx 10^{3.5} \, k_B \text{ K cm}^{-3}$ and $T_0 \approx 10^3$ K.} \autoref{sec:cooling} contains the full details of our cooling and heating source terms. In our fiducial simulation the thermal instability growth timescale is ${\sim}33$ times smaller than the turbulent eddy turnover timescale.\footnote{Note that this value comes from the Milky Way-like cooling and heating and realistic values for the turbulent velocity.} These simulations are, therefore, firmly in the rapidly thermally unstable regime in which we expect a multiphase medium to develop and be sustained despite turbulent mixing. We also compare to simulations where the ratio of the turbulent to thermal instability timescales is reduced to 3 and 1/3.

Our simulations are all initialized with constant pressure $P=P_0$, and density $\rho_0 = \mu m_p P_0 / k_b T_0$ such that the temperature is equal to the unstable equilibrium temperature, $T \equiv \mu m_p P / k_b \rho = T_0$. The perturbations that develop from the driven turbulence quickly cause the medium to separate into a cold and warm phase near the two stable equilibria. The simulations are also initialized with a tangled magnetic field with zero net-flux ($\overline{B} = 0$). As with the velocity field, the magnetic fields are initialized with only large scale perturbations, with equal power in modes with wavenumbers $k L / 2 \pi= 1$ and $2$. Similar initial conditions are commonly used in studies of MHD turbulence \citep[\eg][]{Grete:2021}. Futhermore, it has been shown that in simulations that aim to study turbulent dynamos similar small scale magnetic field structures and the plasmoid instabilities occur as in our simulations that explicitly do not aim to study turbulent dynamos \citep{Galishnikova:2022}. In our simulations the strength of initial tangled magnetic field is set so that the root-mean-square plasma beta is unity, $\beta = \langle P \rangle / \langle B^2/2 \rangle$, as is thought to be the case in the Milky Way ISM \citep{Falgarone:2008}. 

In our 3D simulations we do not use any explicit resistivity, viscosity, or conduction, and instead rely on numerical dissipation. This is necessitated by the prohibitively small length scales required to resolve a small enough explicit resistivity at which the plasmoid instability is captured (relying on numerical resistivity reduces the resolution requirement by a factor of $\sim$5). For comparison, in \autoref{sec:2D} we include a single extremely high resolution 2.5D simulation with explicit, fully-resolved dissipative processes to demonstrate that the behavior seen in the simulations without explicit dissipation are qualitatively consistent, particularly in regards to magnetic reconnection and plasmoid formation. %The simulations with explicit dissipation have $\eta = \nu = \kappa = 10^{-5}$, which corresponds to $Re = 10^5$, $Pr = 1$, and $Pr_M =1$, and, crucially, $S > S_{\rm crit} = 10^4$ for current sheets that are larger that one tenth the total box size. 

Although we prioritize generality with our simulation design and therefore report all results in a dimensionless form, it is useful to relate our unit system to a physical system to guide the readers intuition for the sort of environments our results apply to. Taking the Milky Way ISM as reference in which the average pressure is $P_0 \approx 10^{3.5} \, k_B \text{ K cm}^{-3}$ \citep{Jenkins:2011}, the unstable equilibrium temperature is $T_0 \approx 10^3$ K, and $\Lambda(T_0) \approx 8\times10^{-27}$ \citep{Koyama:2002}, this implies that $\rho_0 \approx 10^{-23.5} \text{g cm}^{-3}$, $n_0 \approx 10^{0.5} \text{cm}^{-3}$, $v_0 \approx 10^{0.5} \text{km/s}$, and $L \approx 75 \text{ pc}$. This can easily be scaled to other systems by adjusting $P_0$, $T_0$, and $\Lambda(T_0)$, and using the fact that $L \propto T_0^{5/2} / (P_0 \Lambda(T_0))$.

Our primary focus is on our highest resolution 3D thermally unstable turbulent MHD simulation that has $\Mach = 1$, $\beta = 1$, and a resolution of $\Delta x = L / 2048$. We also performed several additional simulations with varying parameters, for comparison. To assess convergence in \autoref{sec:converge} we compare to simulations with lower resolutions. In \autoref{sec:2D} we compare 2.5D simulations with only numerical dissipation and with resolved explicit dissipation processes, which we use to confirm the robustness of our findings by showing that plasmoid mediated reconnection is very similar in the two cases. We also compare our fiducial simulation to otherwise identical simulations with either no magnetic fields, no driven turbulence, or an isothermal equation of state. All simulations are analyzed after at least $5 L/c_{\rm s, warm} \approx 5 L/v_{\rm rms, warm}$, which ensures that the turbulence is fully developed and that cooling has had enough time to ensure that the phase structure has reached a steady state. We stress that these simulations are not intended to capture the turbulent dynamo, which takes significantly longer to reach equipartition \citep[\eg][]{Xu:2016, Beattie:2022}. These simulations are at the cusp of what is computationally feasible, so we start our simulations with a near equipartition magnetic field (see \citealt{Galishnikova:2022} for, to our knowledge, the only example of a turbulent dynamo simulation with high enough resolution to resolve the plasmoid instability). In \autoref{sec:longtime} we demonstrate that our results are insensitive to the exact time at which they are measured using a $\Delta x = L / 1024$ simulation that was run for 3.5 times longer ($17.5 L/v_{\rm rms, warm}$). 

%-------------------------------------------%
%-------------------------------------------%
%-------------------------------------------%
%-------------------------------------------%
\section{Numerical Results} \label{sec:results}
We begin with a qualitative investigation of our fiducial, highest resolution thermally unstable magnetized turbulent box simulation, which is followed by a quantitative look at the statistical properties.% of this and our other related simulations. 

\subsection{Magnetic and Thermal Structure on Small and Large Scales}

\begin{figure*}
\centering
\includegraphics[width=0.9\textwidth]{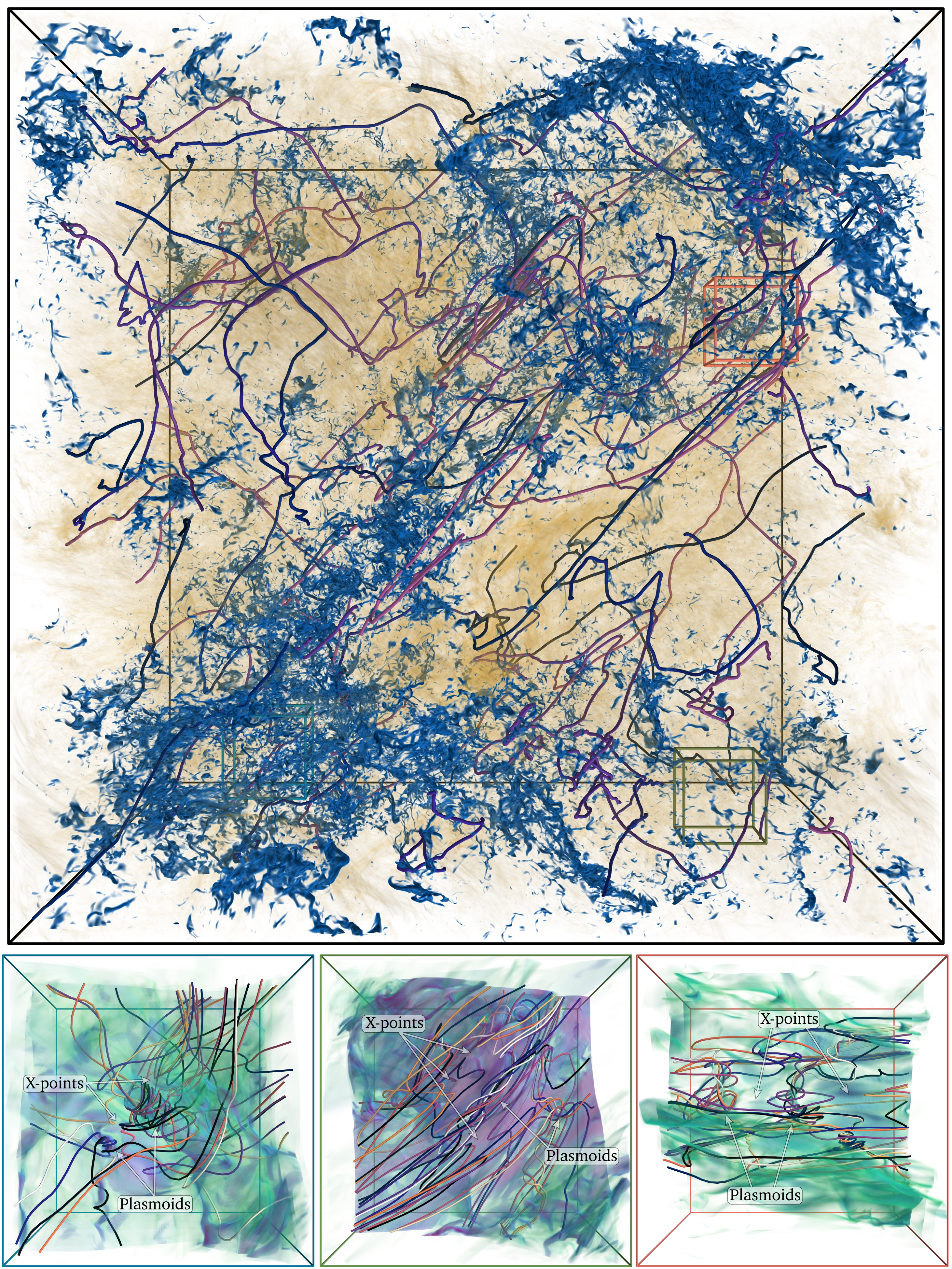}
\caption{(Top) Volume rendering of the temperature (opaque blue corresponds to $T{\approx} T_0 / 100$, and transparent yellow corresponds to $T{\approx} 10 T_0$). Uniformly seeded magnetic field lines are shown in purple---many exhibiting helical patterns that trace plasmoids. (Bottom) Volume rendering of the electric current (increasing from transparent cyan to opaque magenta) and magnetic fields (randomly colored) in three sub-volumes containing plasmoids and ``X-points'' that are $L/8$ on a side. They are shown in full volume as colored cubes (left--blue, center--green, right--red) and respectively exemplify cold, asymmetric, and hot current sheets undergoing plasmoid mediated reconnection. A rotating movie of these can be found \href{https://dfielding14.github.io/movies/}{here}.}
\label{fig:Cubes}
\end{figure*}

\begin{figure*}
    \centering
    \includegraphics[width=\textwidth]{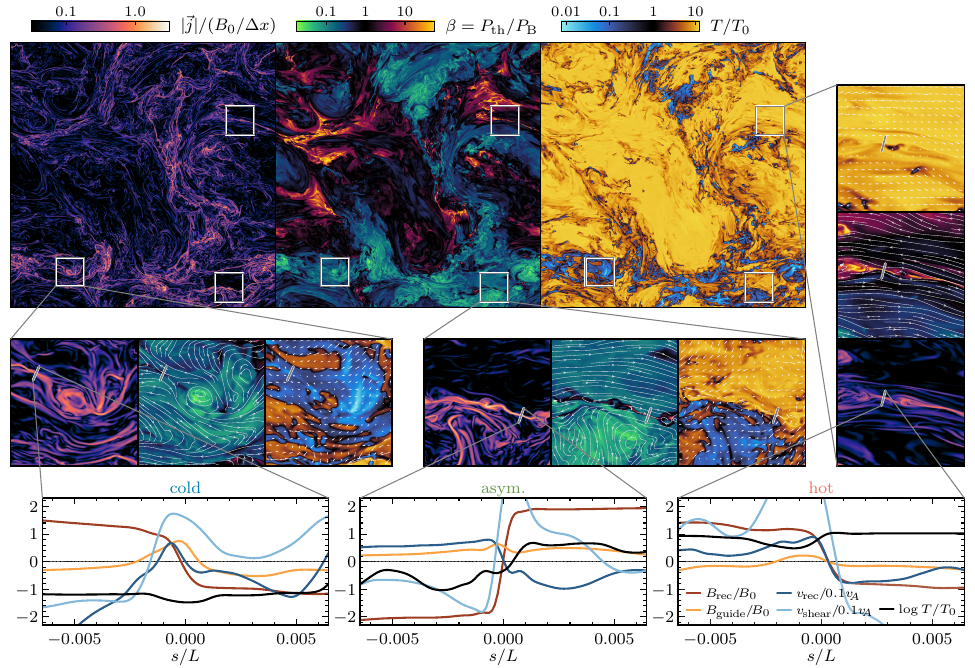}
     \caption{The top panels show 2D slices of the current density, plasma beta ($\beta$), and temperature. Zoom-ins on three regions with plasmoids are shown in the surround panels. In the zoom-in panels the in-plane magnetic fields are shown on top of the $\beta$ map, and the velocity vectors are shown on top of the temperature map. The three zoom in regions are the same as the sub-volumes shown in \autoref{fig:Cubes}, and are in the same order. The zoom-ins are examples of the three different types of regions in which reconnection events that lead to plasmoid formation occur in our simulation: fully cold (left), interface of warm and cold (middle), fully warm (right). For each of the zoom-ins we show profiles perpendicular to the prominent current sheet of the magnetic field components perpendicular ($\Brec$; this is the component that reconnects) and parallel ($\Bguide$) to the local guide field, the velocities into and along the current sheet ($\vrec$ and $\vshear$ respectively) normalized by one tenth the local Alfv\'en velocity ($0.1 \vA$), and the log of the temperature $\log_{10} (T/T_0)$. The current sheets all show a change in direction of $\Brec$, a reconnection speed of $\vrec \sim 0.01 \vA$, and some compression of $\Bguide$ in the cases that have any appreciable guide field. In all cases $\vshear < \vA$, so they should be stable to the Kelvin Helmholtz instability. Animated version \href{https://dfielding14.github.io/movies/}{here}.}
    \label{fig:SliceZoom}
\end{figure*}

In \autoref{fig:Cubes} and \autoref{fig:SliceZoom} we show 3D volume renderings and 2D slices of our fiducial simulation after 5 eddy turnover times of the warm gas ($5 L/v_{\rm rms,warm}$). The top panels of both figures show the full volume, while the surrounding panels show zoom-ins on three regions. These regions (the same in both figures) exemplify the three distinct environments in which we find current sheets undergoing plasmoid-mediated magnetic reconnection: (i) within cold clumps, (ii) at the asymmetric interface of the cold and warm phases, and (iii) within the warm, volume-filling phase. 

The volume on the whole is characterized by a bistable configuration with the vast majority of the gas residing near one of the two thermal equilibria. The cold gas morphology is clearly clumpy with structures on scales ranging from the nearly the size of the whole box ($\sim$10s pc) down to the grid scale (${\sim} 10^{-2}$ pc). Many of these cold gas clumps show an elongated, filamentary structure with large aspect ratios (discussed in \autoref{sec:clumps}). 

The field lines shown in \autoref{fig:Cubes} give a sense for the large and small scale magnetic structure. While some of the volume contains relatively ordered field lines they also generically exhibit highly disturbed regions manifesting as small scale helical knots or ``flux tubes'', which trace plasmoids and are hallmarks of tearing instability induced fast magnetic reconnection. 
This highlights that plasmoids form throughout the volume and can be found in the volume filling warm phase, the clumpy cold phase, and at the interface of the two. 
This is further supported by the 2D slice of the electric current, $\vecj \equiv \curl \vecB$, in the top left panel of \autoref{fig:SliceZoom}. It is clear that there are numerous large current sheets with $L_{\rm CS} \gtrsim L/10$ throughout the volume (a well-known phenomenon in MHD turbulence, see, e.g., \citealt{Zhdankin:2013})---many of which are actively undergoing plasmoid-mediated magnetic reconnection. In \autoref{sec:converge} we show that similar large current sheets form in lower resolution simulations, but they are much more laminar and do not show signs of plasmoid instability. 

The top panels of \autoref{fig:Cubes} and \autoref{fig:SliceZoom} further demonstrate that the thermodynamic and magnetic properties interact non-trivially. Regions with cold gas tend to be more magnetically dominated with $\beta < 1$ (blue-green colors). This is likely due to magnetic field enhancement during cooling induced contraction. Moreover, the regions of highly disordered magnetic fields and large $\vecj$ tend to be associated with thermally unstable or cold gas. %% some sort of punch line here?

Turning now to the zoom-in panels of \autoref{fig:Cubes} and \autoref{fig:SliceZoom} we get a better sense of the nature of reconnection in these environments and the morphology and properties of the plasmoids. Each region has one or more plasmoids\footnote{To our knowledge there is no known robust criterion for identifying plasmoids. We encourage readers to view the animated versions of our plots \href{https://dfielding14.github.io/movies/}{here}, which makes it easier to pick out these structures by eye.} and several clear X-points where the reconnection is actively occurring.\footnote{Relative to the volume renderings of the zoom regions in \autoref{fig:Cubes} the 2D slices of the zoom regions in \autoref{fig:SliceZoom} are oriented into the page horizontally through the middle of the sub-volumes.} Each of these regions exemplify the three distinct environments in which we find evidence for the plasmoid instability. These plasmoids closely resemble the structures seen in local simulations of individual plasmoid unstable current sheets \citep[\eg][]{Daughton:2011, Huang:2016, Yang:2020}, thereby bolstering the evidence that these structures arise from the plasmoid instability in our marginally resolved global simulations. Furthermore, the plasmoid unstable current sheets have a thickness of a few cells (consistent with numerical resistivity and the estimate of \citealt{Grete:2023}) and length of $>200$ cells\footnote{For reference the zoom regions are 256 cells on a side and the example current sheets are wider than the zoom regions. This highlights why resolutions of $\gtrsim 2000$ are necessary since otherwise even the largest current sheets that form, which have $L_{\rm CS} \sim L/8$ would have aspect ratios < 100 and would be stable to the plasmoid instability}. This corresponds to an axis ratio of $\gtrsim 100$ and thus a Lundquist number $S \gtrsim 10^4$, which is comparable to or greater than the threshold for plasmoid instability \citep{Loureiro:2007}. We can, therefore, be confident that these structures arise from the plasmoid instability. We encourage the readers to view the animated versions of our plots \href{https://dfielding14.github.io/movies/}{here} to get a better sense for the morphology and different evolutionary stages of these structures. In general, the plasmoid instability begins in long, relatively undistrubed current sheets, which progress through the linear evolution and develop a series of X-points and plasmoids. The plasmoids eventually evolve into a turbulent nonlinear state.

% In the surrounding panels of \autoref{fig:SliceZoom} we show zoom-ins on three sub-volumes containing current sheets undergoing magnetic reconnection and plasmoid formation. They are highlighted by the boxes on the primary panels. These zoom-in slices are centered on the same sub-volumes as in the bottom panels of \autoref{fig:Cubes}. It is instructive to compare the 3D volume renderings with the 2D slices to get a sense for the complex morphology of these structures and their environs. For reference, the slices shown in \autoref{fig:SliceZoom} are oriented into the page through the middle of the sub-volumes shown in \autoref{fig:Cubes}. In the zoom panels magnetic field lines are plotted on top of the plasma beta $\beta$ slice, and the velocity vectors are plotted on top of the temperature $T$ slice. Beneath each of the zoom-ins we show profiles of key quantities along a ray that passes through the center of the primary current sheet undergoing reconnection. These three regions were chosen because they exemplify the three primary magnetic reconnection and plasmoid formation channels that occur in our simulation.

The zoom region shown on the left is an example of a current sheet that forms in the middle of a cold region. Current sheets like this arise in large part because of the compressive motions that result from the thermally unstable region rapidly contracting in on itself. 
The energy released during reconnection in these cold current sheets is quickly radiated away and the gas inside the plasmoid remains cold. 
This is demonstrated by temperature profile which remains cold along the ray passing through an X-point in the current sheet.

The zoom region shown in the middle is closely related to the first with the important distinction that the current sheet forms at the interface of the warm and cold phase. The pressure and magnetic field strength on either side of the current sheet is---in this example---nearly constant, thus the asymmetry is both in the temperature and density.\footnote{Asymmetric plasmoid unstable current sheets have also been studied in relativistic astrophysical systems, such as the boundaries of jets \citep{Ripperda:2020, Mbarek:2022}.} Perpendicular to the current sheet the temperature changes by nearly a factor of 100 (lower middle panel of \autoref{fig:SliceZoom}). These asymmetric current sheets are formed as a result of the flow of warm gas as it cools onto a cold clump with an oppositely oriented magnetic field. In the ISM, these strong cooling flows, which occur throughout the volume, can lead to the development of plasmoid unstable current sheets. Here reconnection may modify the phase balance as the heating is occurring directly in the thermally unstable phase where it can be most impactful. The plasmoids that form in these asymmetric current sheets tend to be highly elongated, often have cold interiors, and follow the surface of the larger cold clumps.

These first two cases are unique to the thermally unstable ISM-like environments we are considering and represent a new channel for magnetic reconnection. The reconnection is brought on by thermal instability induced tearing instabilities, and leads to the formations of, what we refer to as, \emph{cold plasmoids} (i.e. plasmoids that form as result of radiative cooling and remain cold because of their short cooling times). As we discuss in \autoref{sec:discussion}, cold plasmoids may be readily observable as H\textsc{i} absorption and/or emission, and as sources for strong radio scattering. Furthermore, because of their partial magnetic insulation/shielding these cold plasmoids (as shown in \autoref{fig:Cubes}) can survive for an appreciable fraction of the global dynamical time. In particular, their magnetic structure can make their lifetimes significantly longer than cold clumps of a similar size, which tend be mixed away on an eddy turnover time ($ t_{\rm mix} \sim r_{\rm cl}/v_{\rm turb}(r_{\rm cl}) \lesssim r_{\rm cl}/c_{\rm s, cold} \sim 0.1$ Myr for $r_{\rm cl} = 0.1$ pc). 

The cold and asymmetric plasmoid unstable current sheets can be contrasted to the final zoom region shown in right panels in which a current sheet forms in the volume filling warm phase as a result of the intermittency in the turbulent flow. 
%As the flow drives the current sheet thinner it eventually reconnects, often leading to the formation of plasmoids. 
The current sheet in this case has $\beta \gtrsim 100$, and the energy release from the reconnection has heated the immediate surrounding medium. This \emph{hot plasmoid} formation, as we refer to it since it takes place in the hotter, volume-filling phase, has been seen in simulations of single phase highly magnetized turbulent environments \citep[\eg][]{Chernoglazov:2021,Galishnikova:2022,Dong:2022}, and can be expected to be common throughout the volume filling region of the ISM. 

Looking at the profiles along the rays that pass through the X-point of the most prominent current sheets in each of the zoom regions gives us a more detailed view into the properties of the current sheets. The temperature profiles clearly demonstrate our distinction between the cold, asymmetric, and warm current sheets. Each of the profiles exhibits a large reversal in the reconnecting component of the magnetic field ($\Brec$) over a width of $\sim L/500$. The behavior of the component of the magnetic field aligned with the guide field ($\Bguide$; i.e. the component aligned with the field direction in the middle of the current sheet) is noticeably different between the three profiles. Near the cold current sheet (left) $\Bguide \sim \Brec$, in the asymmetric current sheet $\Bguide \lesssim \Brec / 4$ and remains coherent above and below the current sheet, and in warm current sheet $\Bguide \sim 0$. Therefore, in addition to representing three different thermal properties these three current sheets also represent three guide field properties, namely strong, weak, and zero guide field, respectively. %We leave an investigation of whether the thermal and guide field properties are causally connected to a future work, we suspect, however, that there can be any combination of cold, asymmetric, and warm current sheets that have strong, weak, or zero guide field.

Also shown in the profiles are the inflow velocity $\vrec$ (velocity toward the current sheet that is driving reconnection) and the shear velocity $\vshear$ (velocity along the current sheet) normalized by one-tenth the average local Alfv\'en speed $0.1 \vA$. In all cases $\vrec{\sim}0.01{-}0.03 \vA$. This reconnection rate is comparable to the fast reconnection expected for tearing mediated reconnection \citep[\eg][]{Loureiro:2007,Bhattacharjee:2009,Uzdensky:2010}. Furthermore, in all cases $\vshear < \vA$, which implies that the current sheets are stable to the Kelvin Helmholtz instability \citep{Loureiro:2013,Chernoglazov:2021}. We further investigate the impact of rotational flows on current sheets in \autoref{sec:vorticity}, and find that some plasmoids form in regions with large vorticity, which may be due to a stabilizing effect from the transverse component of the magnetic fields \citep{Somov:1989}. That said, it is likely that some of the plasmoid/flux tube structures in our simulation arise as a result of instabilities other than the plasmoid instability. This only bolsters our result that these structures are likely present in the ISM. %In the asymmetric current sheet $\vrec$ is somewhat larger than the expected 0.01 $\vA$, which we ascribe to the additional driving due to the thermal instability induced contraction.

\subsection{Statistical Properties} 

We now shift our attention to the statistical properties of the simulation with an eye towards understanding the combined effects of turbulence, thermal instabilities, and magnetic reconnection.

\subsubsection{Magneto-Thermal Phase Structure}

\begin{figure*}
\centering
\includegraphics[width=\textwidth]{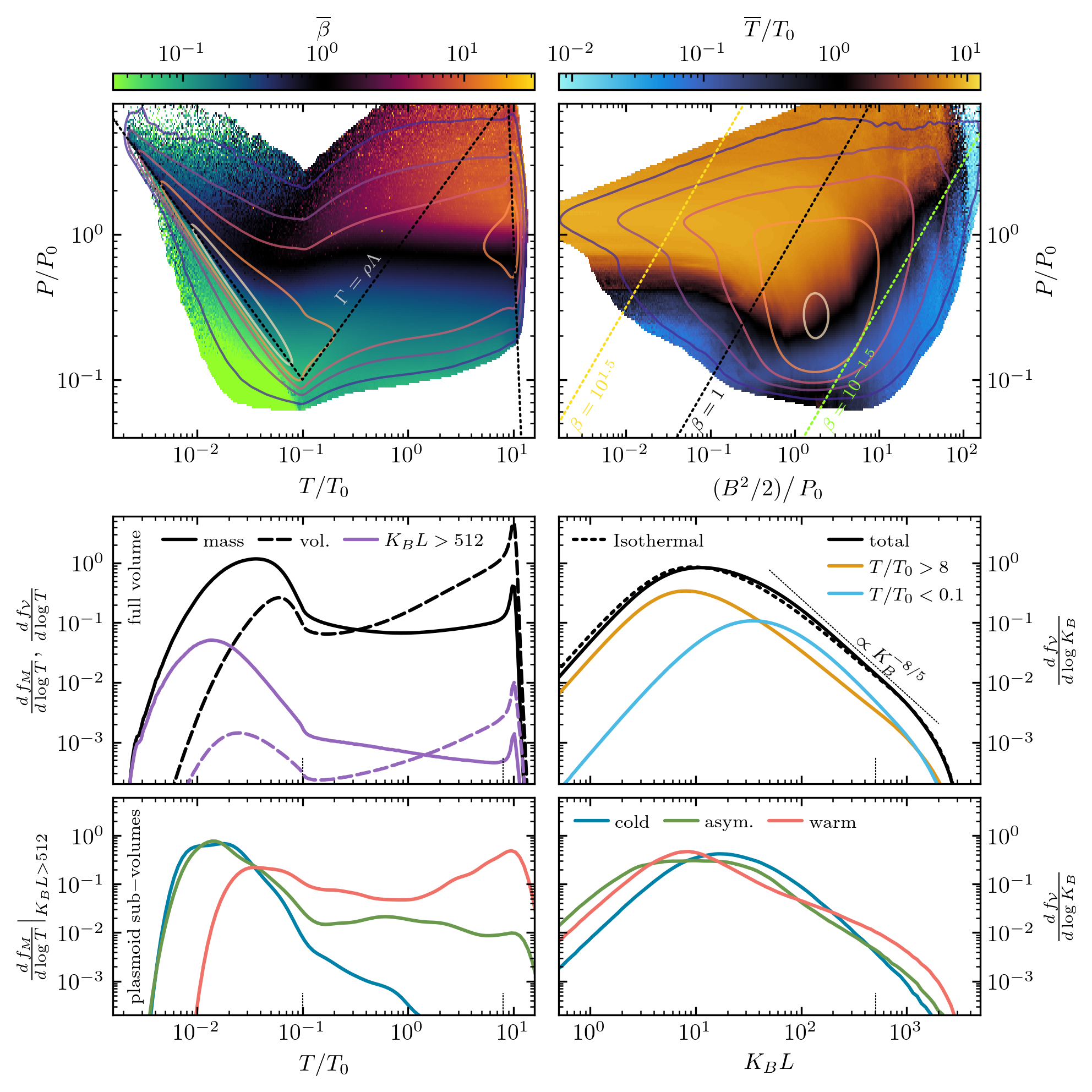}
 \caption{Summary of the magneto-thermal phase structure. The top left and right panels show the joint thermal pressure-temperature phase distribution colored by the average plasma beta ($\beta$), and thermal pressure-magnetic pressure phase distribution colored by the average temperature, respectively. The contours trace the fraction of the total mass per logarithmic bins spaced by 1 dex from $10^{-4}$ to 1 going from purple to white. The thermal equilibrium curve ($\Gamma = \rho \Lambda$) is shown in the black dotted line in the top left panel, and lines of constant $\beta$ are shown in the colored dotted lines in the top right panel. The mass ($f_M$; solid lines) and volume ($f_\mathcal{V}$; dashed lines) weighted temperature distributions are shown in the middle-left panel for the full volume (black lines) and for the high curvature regions only ($K_B > 512/L$; purple lines). The mass weighted temperature distributions of the high curvature regions in the example plasmoid sub-volumes (same as in \autoref{fig:Cubes} and \autoref{fig:SliceZoom}) are shown in the lower left panel. The volume distributions of magnetic curvature of the total volume (black), cold gas (blue), and warm gas (yellow) are shown in the middle-right panel. The dotted line shows the curvature distribution for the isothermal simulation. The curvature distributions in the plasmoid sub-volumes are shown in the lower right panel.}
\label{fig:PDF}
\end{figure*}

\autoref{fig:PDF} provides an overview of the thermal and magnetic properties, and their correlations. 
We unpack these results incrementally starting with a discussion of the middle-left panel, which shows the temperature distribution of the mass (solid lines) and volume (dashed lines). The total mass of the domain is dominated by a broad distribution of cold gas with $T<T_0 / 10$. There is a peak of mass at the warm phase equilibrium near $10 T_0$. Connecting these two peaks, and populating the thermally unstable regime, is a nearly flat distribution, that is continually depleted and repopulated by the combined effects of radiative cooling and heating, turbulent mixing, and reconnection heating. The volume, on the other hand, is dominated by the warm phase. This is a reflection of multiphase nature shown in the previous figures. 

%% I could probably delete this whole paragraph --- or just keep the sentence that slower cooling = less cold phase
What determines the exact balance between the amount of mass in each of the two phases remains an open question. The answer depends on a non-trivial combination of the primary physical processes at play. The strength of conduction has been shown to regulate the balance in more simplified systems \citep[\eg][]{Zel'Dovich:1969,Choi:2012,JGKim:2013,Jennings:2021}. Furthermore, we found in a simulation with a longer cooling time relative to the eddy turnover time (e.g., a smaller box), the fraction in the cold phase decreased. Likewise, increasing the strength of turbulence can significantly modify the phase distribution and in the extreme limits can erase the bimodal structure altogether \citep[\eg][]{Vazquez-Semadeni:2000, Gronke:2022}. How magnetic reconnection may impact this balance is an important future consideration given the difficulty of adequately resolving this process in ISM simulations. 

To partially illuminate both the role of magnetic reconnection on the thermal structure and the role of thermal instabilities on the magnetic structure we also show with the purple lines in the middle-left panel of \autoref{fig:PDF} the mass and volume temperature distributions of gas with large magnetic curvatures, $K_B~\equiv~(\hat{b}~\cdot~\boldsymbol{\nabla})~\hat{b}$, where $\hat{b} = \vecB/|\vecB|$. 
%The magnetic curvature is a useful metric for how rapidly the magnetic field changes direction along its length and is inversely proportional to the field reversal length scale. 
We adopt a threshold of $K_B > 512/L$ which corresponds to the field changing direction in four cells or less (the center-right panel shows the full $K_B$ distribution and the small dashed line indicates our choice of threshold). 
High $K_B$ is good indicator of magnetic reconnection because of the large curvature inherent to plasmoids and X-points (as can be seen in the lower-right panel, which shows the curvature distribution of the plasmoid sub-volumes). The temperature distributions of the high $K_B$ gas are skewed toward more cold gas than the volume as a whole. This indicates that current sheet formation and reconnection is enhanced in the cold phase---likely as a result of the compression of the field due thermal instability induced contraction. 

The curvature, $K_B$, alone is an imperfect tracer of plasmoids, so in the lower-left panel of \autoref{fig:PDF} we show the temperature distribution of the high $K_B$ gas in the three example sub-volumes that contain clear plasmoids shown in the previous figures. All three sub-volumes exhibit a multiphase structure, however the cold current sheet has a negligible amount of warm, highly magnetically curved material. This demonstrates a key point that the heating by magnetic reconnection can be rapidly lost due to radiative cooling and is thus insufficient to drive the material to the hotter phase. How this heating impacts the global phase structure is an interesting discussion for future works.

The volume distribution of $K_B$ in the whole volume (black), cold gas (blue), warm gas (yellow), and in the example plasmoid sub-volumes is shown in the middle and lower right panels of \autoref{fig:PDF}. As expected the plasmoid sub-volumes have a large high curvature volume fraction owing to the fact that plasmoids are, by definition, comprised of highly curved field lines. At the high curvature end the total curvature distribution scales proportional to ${\propto} K_B^{-8/5}$. On average magnetic fields in cold gas ($T<T_0/10$) tend to be more curved than in the warm gas ($T> 8 T_0$), but the highest $K_B$ values are equally likely in either phase. The curvature distribution of all of the gas is remarkably similar to the isothermal driven turbulence simulation we ran for comparison. 
%Nevertheless it is clear that there are important correlations between the magnetic and thermal properties of the gas.

To better understand the magneto-thermal phase distributions, in the the top left panel of \autoref{fig:PDF}, we show the joint distribution of pressure and temperature (contours trace the mass distribution and colors show the average $\beta$). As expected, the majority of the mass is found near the two stable thermal equilibria (i.e., along the negatively sloping portions of the equilibrium curve). The gas near the warm equilibrium is tightly clustered around $P \approx P_0$ and $T \approx 10 T_0$, and has an average $\beta \sim 1$. The mass near the cold equilibrium is spread out along the stable equilibrium curve with $10^{-2.5} \lesssim T/T_0 < 0.1$ and $0.1 \lesssim P/P_0 < 3$. The $\beta$ also varies in the cold phase, with the lower pressure cold gas having $\beta \lesssim 1/10$---indicating a strong compensation of the thermal pressure deficit by enhanced magnetic pressure ($B^2 /2$). The enhancement of magnetic pressure in regions with lower thermal pressure means that, although the gas near the cold and warm thermal equilibria are not in thermal pressure equilibrium, they are close to being in total pressure equilibrium with $P + B^2 /2 \approx \text{constant}$. There is some mass with very low temperatures $T \lesssim T_0 / 100$ that has both high thermal pressure and low $\beta$, which is found in rapidly contracting thermally unstable regions that often host cold current sheets. 

The upper right panel of \autoref{fig:PDF} shows the joint mass distribution of thermal pressure and magnetic pressure and is colored by the average temperature in each bin. Most of the mass is found at a pressure of $P\approx0.3 P_0$ and a magnetic pressure of $B^2 / 2 \approx 2 P_0$, corresponding to $\beta \approx 10^{-0.75}$. The mass distribution has a distinct shape with a tail towards extremely low magnetic pressures. This tail follows a line of nearly constant total pressure. Gas moves \emph{both} ways along lines of nearly constant total pressure. On the one hand, as warm gas cools and contracts it loses thermal pressure and gains a nearly equal proportion of magnetic pressure. On the other hand, highly magnetized gas that undergoes reconnection moves in the other direction in this phase space---the conversion of magnetic energy into thermal energy raises the thermal pressure of the gas as it loses magnetic pressure. In some cases, however, the magnetic reconnection heating is not sufficient to overcome the strong radiative cooling, so the end result is cold gas with low thermal pressure and low magnetic pressure, as can be seen by the lobe in the mass distribution of cold gas with $P\sim 0.3 P_0$, and $B^2 / 2 \lesssim 10^{-1}$. This is the case in the \emph{cold plasmoid} formation channel exemplified by the cold and asymmetric current sheets shown in the previous two figures and in the bottom panels of \autoref{fig:PDF}. Note that in addition to heating, reconnection also accelerates gas, so some appreciable fraction of the magnetic energy is also being converted to small scale motions.

\begin{figure*}
\centering
\includegraphics[width=\textwidth]{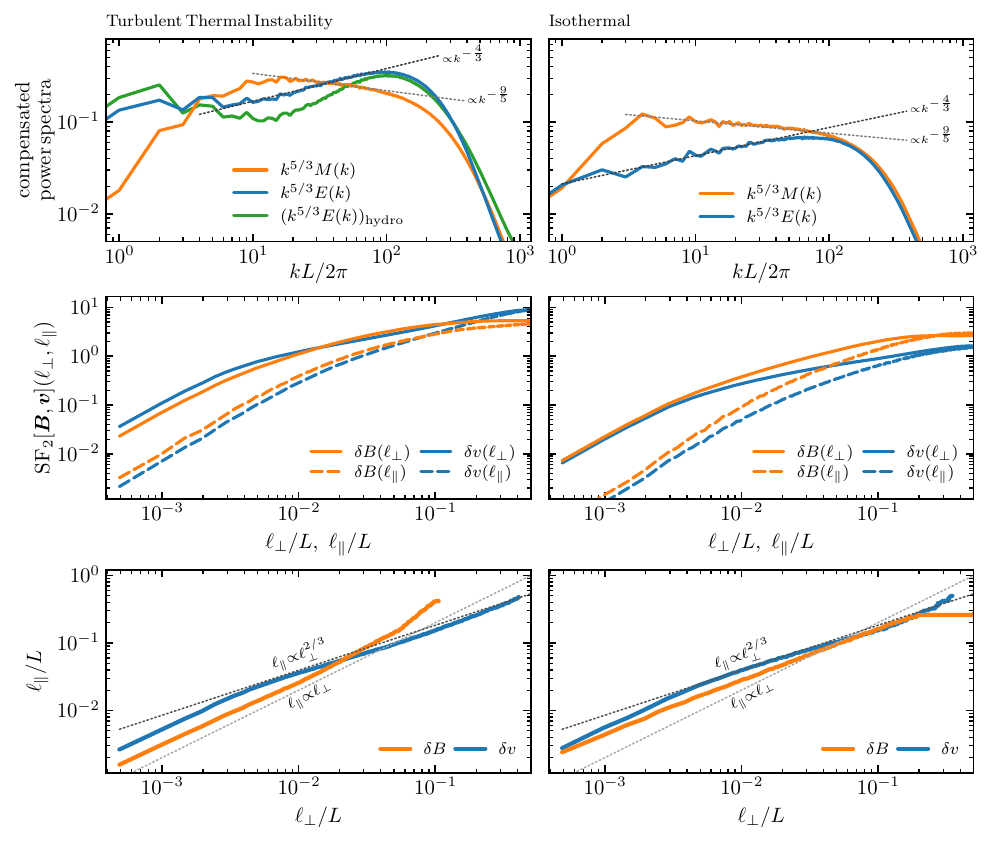}
 \caption{Statistical measures of the scale-dependent magnetic field and velocity structure. The left column shows the results from our turbulent thermal instability simulations and the right column shows the results from the isothermal turbulence simulation. The top panels show the magnetic $M(k)$ (orange) and velocity $E(k)$ (blue) power spectra compensated by $k^{5/3}$. Approximate power-law fits are shown in the thin dotted lines. Also shown in the upper left panel is the velocity spectrum from the non-magnetized simulation (green). The middle row shows the second order magnetic and velocity structure functions parallel and perpendicular to the local mean field. The bottom row shows the anisotropy of the magnetic and velocity fluctuations defined to be the scale at which the parallel and perpendicular structure function are equal.}
\label{fig:PS_SF}
\end{figure*}

The detailed analyses presented in \autoref{fig:PDF} highlights the rich interplay between the thermodynamics, turbulence, and magnetic fields that combine to regulate the magneto-thermal phase structure of the ISM. While there is clearly still much to be understood (particularly in the computationally daunting limit of resolved explicit resistivity, conduction, and viscosity), we have shown that there are important correlated magnetic and thermal properties that when simulated at high enough resolution can lead to the formation of cold and hot plasmoids.

\subsubsection{Scale Dependent Magnetic and Kinetic Structure}

We now move on to an investigation of the structure of the magnetic and velocity fields as a function of scale, which is summarized in \autoref{fig:PS_SF}. The top row shows the magnetic ($M(k)\equiv \int d\Omega_k k^2 \langle |\vecB(\pmb{k})|^2 \rangle/2$) and velocity ($E(k)\equiv \int d\Omega_k k^2 \langle |\vecv(\pmb{k})|^2 \rangle/2$) power spectra compensated by $k^{5/3}$ for the turbulent thermal instability (left; $E(k)$ of a non-magnetized simulation shown in green) and isothermal driven turbulence simulations (right). 
On large scales (small $k$) in the turbulent thermal instability simulation $E(k) > M(k)$, but $M(k)$ is nearly flat while $E(k) \propto k^{-5/3}$, which leads to a reversal at intermediate scales ($4 \lesssim k L/2\pi \lesssim 32$), in which $E(k) < M(k)$. Starting on intermediate scales there is a broad range of $k$ over which the $M(k)$ steepens to approximately follow $\propto k^{-9/5}$, and the $E(k)$ flattens to follow $\propto k^{-4/3}$. This transition and change in slopes occurs at $k L / 2 \pi \approx12$. On small scales ($k L / 2\pi \gtrsim 32$), $E(k)$ again dominates $M(k)$. These general trends and slopes are well converged with numerical resolution down to 16 times lower resolution ($\Delta x = L/128$) than our fiducial simulation ($\Delta x = L/2048$; see \autoref{sec:converge}). Most of these same trends were found in the similar (albeit significantly lower resolution) turbulent thermally unstable MHD simulations presented by \cite{Kritsuk:2017}, with the notable exception of the dominance of $E(k)$ at large $k$, which we ascribe to slower cooling relative to turbulent mixing in their simulations.
The slope of the power spectra in simulations with additional physics, such as those presented in \citealt{Padoan:2016} that include self-gravity and SNe, exhibit somewhat steeper $E(k)$, and flatter $M(k)$.
We remind the readers that these simulations are highly idealized and are not attempting to reproduce specific observations, which tend to show somewhat different slopes than what arises in our simulations \citep[\eg][]{Chepurnov:1998,Lee:2020}. Furthermore, it is important to note that the shape of the energy spectra near the end of the inertial range is subject to some distortion from the bottleneck effect \citep[\eg][]{Schmidt:2004}.

In contrast, in the isothermal simulation the $M(k) > E(k)$ on all scales, although they do approach equipartition on the largest scales ($kL/2\pi \sim 1$) and on the smallest scales ($kL/2\pi \lesssim L / 16 \Delta x$). The velocity spectrum follows a shallower $\propto k^{-4/3}$ from $k L/2 \pi \sim 1$ down to $\sim 100$, which is in agreement with other similar simulations \citep[\eg][]{Grete:2021}. The magnetic spectrum, on the other hand, peaks around $k L/2\pi \sim 4$ and then follows a relatively steep $\propto k^{-9/5}$ slope down to the dissipative scale. These slopes, which are robust to changes in resolution, agree very well with the slopes of the thermally unstable simulations at $k L/2\pi \gtrsim 12$.

Comparing the spectra of the thermally unstable and isothermal simulations highlights the enhanced kinetic energy at small scales in the thermally unstable simulations. This cannot be due to magnetic reconnection alone since both exhibit plasmoid-mediated magnetic reconnection. Furthermore, with our resolution, $\Delta x = L / 2048$, only the largest current sheets are resolved and the scales on which plasmoids are forming are somewhat impacted by numerical dissipation. We, therefore, cannot say that at yet higher resolution that plasmoid-mediated magnetic reconnection would or would not change the power spectra as is predicted for a turbulence with a strong mean field \citep{Boldyrev:2017,Dong:2022}. 
Comparing to the spectrum of a non-magnetized turbulent thermally unstable simulation, which also exhibits a similar large $k$ peak in $E(k)$ (green line in top left panel of \autoref{fig:PS_SF}), supports the notion that strong radiative cooling is driving small scale motions in the thermally unstable simulations. Further evidence supporting this picture comes from the fact that as we reduced the strength of cooling in our simulations (not shown) the magnitude of the small scale (large $k$) peak in the velocity spectra also decreased. The spectra in both the magnetized and non-magnetized turbulent thermally unstable simulations change slopes at $k \sim 10-20$, which is a reflection of the imprint of cooling on the flows.

%The energy spectra are inherently isotropic and so cannot provide any information on the degree to which velocity and magnetic field perturbations are aligned with the local mean magnetic field. 
To further understand the structure of the magnetic and turbulent velocity fields, we turn to the structure functions shown in the second row of \autoref{fig:PS_SF}. They encode the scale-dependent structure and anisotropy with respect to the local magnetic-field direction. We adopt the two-point second order structure function that is defined for the magnetic field to be ${\rm SF}_2 [\vecB](\vecl) \equiv \langle  |\vecB(\pmb{x}) - \vecB(\pmb{x} + \vecl)|^2  \rangle$, and equivalently for the velocity field. In a sufficiently magnetized plasma these structure functions can show a dependence on the degree of alignment between the separation vector $\ell$, and the local mean magnetic field, which we define to be $\overline{\vecB}_\ell \equiv (\vecB(\pmb{x}) + \vecB(\pmb{x} + \vecl))/2$ \citep{Cho:2000}. We, therefore, consider the parallel and perpendicular structure functions, which are defined to be ${\rm SF}_2 [\vecB](\lpar) \equiv \langle  |\vecB(\pmb{x}) {-} \vecB(\pmb{x} {+} \vecl)|^2; 0 {\leq} \theta_\ell {<} \tfrac{\pi}{36}  \rangle$, and ${\rm SF}_2 [\vecB](\lperp) \equiv \langle  |\vecB(\pmb{x}) {-} \vecB(\pmb{x} {+} \vecl)|^2; \tfrac{17\pi}{36} {\leq} \theta_\ell {<}\tfrac{\pi}{2}  \rangle$ where $\theta_\ell \equiv \arccos |\overline{\vecB}_\ell {\cdot} \vecl / \overline{B}_\ell \ell|$ is the angle between the displacement vector and the mean magnetic field direction \citep[cf.][]{St-Onge:2020}. The average difference in magnetic field and velocity at a given distance perpendicular to the local mean magnetic field is significantly smaller than the difference the same distance parallel to the local mean magnetic field in both the thermally unstable and isothermal simulations. This is inline with the notion that it is easier to perturb the field and the flow perpendicular to the local mean magnetic field than it is parallel. %The perpendicular and parallel structure functions of both the velocity and magnetic field approximately follow ${\propto}\ell^{1/2}$ and ${\propto} \ell^{5/6}$ scalings, respectively. 

\begin{figure*}
\centering
\includegraphics[width=\textwidth]{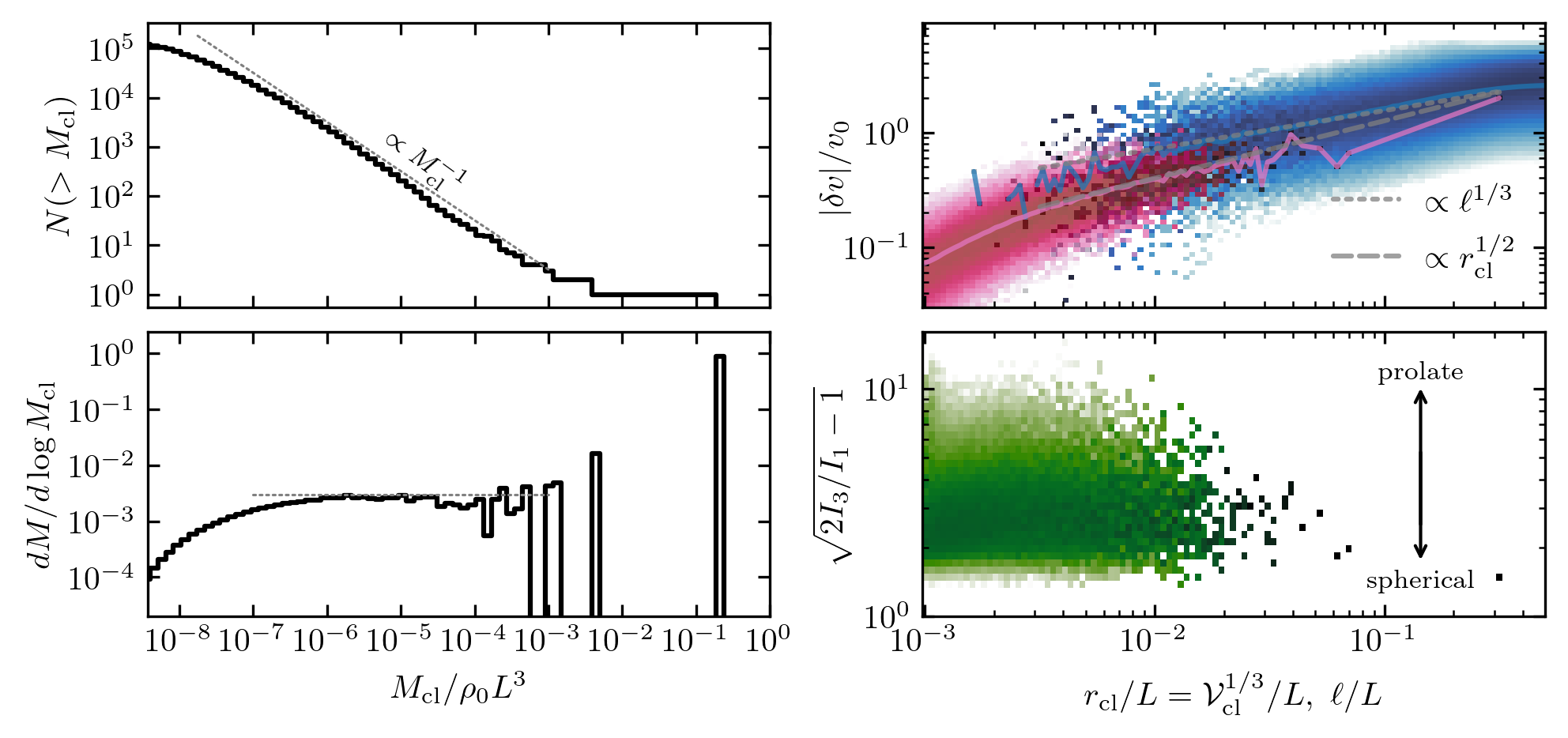}
\caption{Distribution and properties of the cold clumps in our fiducial turbulent thermally unstable simulation. Clumps are defined to be contiguous regions of material on the cold stable thermal equilibrium. The left panels show the distribution of clump masses. The upper right panel shows the internal velocity dispersion as a function of clump size (given by the cube root of the clump volume $r_{\rm cl} = \mathcal{V}_{\rm cl}$) in pink, and the pairwise velocity difference of the clumps as a function of separation ($\ell$) in blue. The solid lines show the median values. The lower right panel shows the clump elongation versus clump size with larger values indicating highly elongated clumps, and values $\sim 1-2$ indicating nearly spherical objects.}
\label{fig:Clumps}
\end{figure*}

We can take this one step farther by investigating the scale dependent anisotropy of both the magnetic field and velocity fluctuations, which are shown in the bottom row of \autoref{fig:PS_SF}. This is calculated by equating the parallel and perpendicular structure functions, which yields the parallel length scale $\lpar$ as a function of the perpendicular length scale $\lperp$ (or vice versa) for both the magnetic fields ($\ell_{\parallel,\delta B}$) and the velocity ($\ell_{\parallel, \delta v}$) fluctuations \citep[\eg][]{Cho:2000,Maron:2001}. In the isothermal simulation when $\lperp \gtrsim 8 \Delta x$ the magnetic field and velocity fluctuations are increasingly elongated along the local mean magnetic field with decreasing scale following the predicted $\ell_{\parallel} \propto \ell_{\perp}^{2/3}$ \citep{Goldreich:1995,Lazarian:1999}. On the other hand, in the thermally unstable simulation only the velocity fluctuations follow this scaling, while the magnetic fluctuations instead have $\ell_{\parallel,\delta B}$ scaling linearly or even super-linearly with $\ell_{\perp,\delta B}$. This change in the anisotropy scaling must come about due to the cooling inducing flows that are sufficiently strong to rearrange the field. Note that there is a transition to isotropy at small scales due to numerical dissipation. 
%This is supported by the results of a similar thermal instability simulation we performed (with four times lower resolution) without any turbulent driving (only small perturbations in the initial conditions to seed the growth of the instability) that exhibited the same magnetic fluctuation isotropy and velocity fluctuation anisotropy.% (see \autoref{sec:no drive}). 

\subsection{Clump Mass Distributions}\label{sec:clumps}

We end this section with a brief look at the properties of the cold clumps that form in our fiducial turbulent thermally unstable simulation. We identify cold clumps in the simulation using a pressure-temperature cut that selects gas that is close to the cold thermal equilibrium\footnote{Specifically the threshold we adopt is $P T < 1.001 T_{\rm cold}$ and $T < 10 T_{\rm cold}$. This picks out the gas sitting near the cold thermal equilibrium. We discard all clumps comprised of 8 cells or less.} (see top left panel of \autoref{fig:PDF}).  The top left panel of \autoref{fig:Clumps} shows clump mass, $M_{\rm cl}$, distribution, and the bottom left panel shows the amount of mass contained in cold clumps per logarithmic bin in clump mass. Over nearly 4 orders of magnitude in clump masses the distribution closely follows a Zipf's Law like distribution with an index of -1, which has commonly been observed in other turbulent multiphase systems \citep[see][for a recent investigation of the origin of this distribution]{Gronke:2022}. This distribution yields nearly equal mass per logarithmic bin in clump mass. In terms of total mass, however, the distribution is dominated by a single large clump that contains $> 30$ percent of all the mass in the domain. The amount of mass contained in this single large cloud is dependent on the strength of cooling. In similar simulations with weaker cooling relative to turbulent mixing, the amount of mass in the largest cloud is reduced dramatically, while the slope of the main distribution is nearly unchanged.

In the top right panel of \autoref{fig:Clumps} we show the intra- and inter-clump velocity distributions as a function of clump size and separation, respectively. The intra-clump velocity (defined internal velocity standard deviation $v_{\rm cl, std}$) versus clump size distribution is shown in the pink along with the median relation. The distribution closely follows a Larson's Law like scaling with $v_{\rm cl, std} \propto r_{\rm cl}^{1/2}$ \citep{Larson:1981}. This is also consistent with the scaling for supersonic Burgers turbulence, as expected since $v_{\rm cl, std} > c_{\rm s, cl} \approx 0.1 v_0$ for all clumps. The inter-clump velocity (difference between two clumps; the mean of which is the first order clump velocity structure function) follows a somewhat flatter relation with $\langle | \delta v_{\rm cl}(\ell)| \rangle \propto \ell^{1/3}$, which follows the prediction for Kolmogorov-like turbulence and is consistent with recent observations \citep{Ha:2022}. The difference between these two scalings points to the cold clouds being entrained in a background sub/trans-sonic turbulent flow while having supersonic internal motions. 

%In the bottom right panel \autoref{fig:Clumps} we show the distribution of the magnetic field standard deviation to mean ratio within a clump versus clump size. The mean of the distribution is shown in the black line, which closely follows a linear relation $\delta B_{\rm cl} / \overline B_{\rm cl} \propto r_{\rm cl}$. This demonstrates that the well known trend that on small scales the field tends to be more aligned than on large scales holds within cold clumps themselves. 

In the bottom right panel \autoref{fig:Clumps} we show the elongation distribution as a function of clump size. To quantify the elongation we adopt $(2 I_3 / I_1 -1)^{1/2}$ as a metric where $I_3$ and $I_1$ are the largest and smallest moments of inertia of the clump, which for a prolate (cigar shaped) ellipsoid equals the ratio of the length of the longest axis to the shortest. For a sphere this takes a value of 1, and for an thin disk this reaches $5^{1/2}$. Although this quantity is imperfect for distinguishing between marginal cases it clearly demonstrates that the cold clumps in our simulation tend to be highly elongated structures. 

\section{Discussion and Conclusion} \label{sec:discussion}
% Summary
\subsection{Summary}
We present the highest resolution (to our knowledge) magnetohydrodynamic simulation of a thermally unstable turbulent medium to date.
We show that large current sheets unstable to plasmoid-mediated reconnection form regularly throughout the volume (the simulations in previous related studies, such as \citealt{Padoan:2016,Kritsuk:2017}, had significantly lower resolution, which prevented them from resolving the critical Lundquist number, $S_{\rm crit}$, and, therefore, inhibited the plasmoid instability).
The current sheets that reconnect and give rise to plasmoids that develop in three distinct environments: within the coldest phase, at the asymmetric interface of the cold and warm phases, and within the volume-filling warm phase. 

We show the rich magneto-thermal phase structure that develops, in which that most of the mass is in the highly magnetized cold phase with $\beta < 1$ (albeit super-Alfv\'enic, $\MachA > 1$), and that the regions of highest magnetic curvature span a broad range in temperatures. Furthermore, we show that thermal instabilities and magnetic fields change the scale dependent magnetic and velocity structure. This is achieved by comparing our fiducial simulation to an isothermal simulation and a simulation without magnetic fields. The comparison highlights that the velocity spectrum dominates on both large and small scales with cooling likely being responsible for the increase in small scale kinetic energy. Additionally, in the isothermal simulation both the magnetic and velocity perturbations follow the predicted increase in anisotropy relative to the local mean magnetic field with decreasing scale. In the thermally unstable simulation, however, only the velocity perturbations follow this scaling whereas the magnetic perturbations are more isotropic and maintain the same small degree of anisotropy on all scales. 

Lastly, we show that the clumps of cold gas in the thermally unstable simulation mostly follow a scale free mass distribution with equal mass per logarithmic bin in clump mass, with the exception of the largest clump that contains most of the mass in the entire simulation. These clumps tend to be highly elongated and exhibit a size versus internal velocity relation consistent with supersonic turbulence, and a relative clump distance-velocity scaling consistent with subsonic motion. 

%Caveats
\subsection{Known Limitations}
While our simulations presents a major step forward in resolving the small scale magneto-thermal properties of the ISM they are clearly, and intentionally, highly idealized, which means there are important caveats that should be taken into account. 
Although our piecewise power law cooling and heating source terms capture the important general trends applicable to a range of systems, they miss important subtleties essential for detailed comparison to specific systems. 
Likewise our simulations do not track changes the chemistry and ionization state of the system, which can have important observational and physical impacts. 
For our 3D simulations, we rely on numerical dissipation to break the magnetic flux freezing when in reality explicit non-ideal MHD effects, such as ohmic dissipation, should be mediating the reconnection. Resolving these effects in the plasmoid-mediated regime in 3D is extremely computationally expensive and has only been achieved once in a single phase, decaying turbulence simulation \citep{Dong:2022}. We stress, however, that in \autoref{sec:2D} we use 2.5D simulations to test the impact of using resolved explicit dissipation and find strong indications to believe that the physical processes we see are robust. 
Similarly, we do not include explicit conduction and viscosity in our 3D simulations which is expected to be relevant \citep[\eg][]{Cowie:1977} and highly anisotropic \citep[\eg][]{Choi:2012}, and may alter the phase structure and flows in important ways. 
Supernovae, which we have not included in our simulations, generate a third, hottest phase \citep{McKeeOstriker:1977} and drive more realistic turbulence than what we assumed. It remains to be seen how reconnection proceeds in the hot phase, which will be much more subsonic, and in the presence of more compressive flows.
%Of particular interest is how the third, hottest phase of the ISM, which we have not included in our simulations and is generated by supernovae , will modify or enhance our new view of plasmoid-mediate magnetic reconnection in the ISM. Supernovae are a major source of turbulence driving in the ISM (along with galactic shear) and may provide different mechanisms for plasmoid-unstable current sheets to form. 
And finally, our simulations do not include the impact of gravity, which may modify the dynamics particularly in the densest regions and may lead to the development of converging flows that can form current sheets. Despite these caveats our findings are relevant to many outstanding ISM observation puzzles and may provide a useful guide to better understanding the physical processes at play in the ISM. 
% move to discussion
% We end this subsection by stressing that although these simulations are very high resolution it is only the largest the current sheets that are resolved well enough to be tearing unstable, and those only marginally. Given the exceptionally small predicted resistivity of the ISM (and many other astrophysical plasmas), we, therefore, expect that in the limit of infinite resolution plasmoid mediated magnetic reconnection would be common down to extremely small scales, orders of magnitude smaller than we can resolve here. There is good reason to believe that the three types of tearing unstable current sheets we have identified here---cold, asymmetric, and warm---would continue to be present on vastly smaller scales. 

%Observations
\subsection{Observational Implications}
In the interstellar medium it is well known that structures exist across a vast range of scales spanning more than 10 orders of magnitude from hundreds of pc down to hundreds of km \citep{Armstrong:1995}.
On the few to ${\sim }10^4$ AU scale there is ample observational evidence for tiny scale atomic and ionized structures (TSAS and TSIS, respectively) that probe the poorly understood dissipation processes \citep[for a useful overview, see][]{Stanimirovic:2018}. These structures manifest as extremely small or fast spatial or temporal variations in H\textsc{i} absorption (TSAS) and as extreme scattering events and interstellar scintillation in radio observations of both quasars and pulsars (TSIS). Observational models favor either very high density spherical structures or highly elongated structures viewed lengthwise. The plasmoids in our simulation are both dense \emph{and} elongated, and they are found on the smallest scales we are able to probe which is comparable to the larger end of the observed sizes ($\Delta x \sim 5\times10^{3}$ AU). Plasmoids will only become more common at smaller scales if we were able to probe higher resolutions. The thermal instability induced cold plasmoids are thus good candidates for TSAS, and the hot plasmoids are good candidates for TSIS. A careful comparison will need to be made to show this connection more concretely, but the qualitative agreement is compelling. Furthermore, the magnetic topology of the plasmoids provides a natural confinement and insulation which will tend to extend the lifetime of these small structures. Our scale separation between the outer and dissipative scales is, however, limited, so a better understanding of how these results scale is needed.

On somewhat larger scales recent work has uncovered highly elongated thin H\textsc{i} structures referred to as H\textsc{i} fibers \citep{Clark:2014}. These fibers are dense ($n \sim 14 \,{\rm cm}^{-3}$), have lengths on the order of a few pc and widths $\lesssim 0.1$ pc (possibly $\ll 0.1$ pc since the widths are unresolved in current observations). The properties of the cold plasmoids in our simulations are highly reminiscent of the properties of the H\textsc{i} fibers and may provide a physical explanation for their origin. This supports a physical picture in which TSAS and H\textsc{i} fibers are cold plasmoids seen either length-wise or edge-wise. A detailed look (which we defer to a future work) at the properties of the cold clumps in our simulations, and in particular how aligned they are with the local magnetic field may help shed light on the nature of ISM turbulence \citep[\eg][]{Gonzalez-Casanova:2017} and far ranging observations such as foregrounds in the Cosmic Microwave Background \citep{Clark:2021}.

\subsection{Physical Implications}
%Cosmic Rays and anisotropic conduction
Our findings have potentially major implications for the transport of cosmic rays (CRs) in the ISM. CRs play an essential role in the evolution of galaxies \citep{Zweibel:2017}, but their impact depends strongly on their transport \citep{Salem:2014,Ruszkowski:2017,Hopkins:2020,Quataert:2022b,Quataert:2022}.
CR transport is regulated by the scattering with small-scale electromagnetic fluctuations that are either self-excited \citep{Kulsrud:1969,Skilling:1971}, preexist in the ambient turbulent medium \citep[\eg][]{Chandran:2000,Yan:2004}, or some combination of both \citep{Xu:2022}.
Recent work has shown that both of these models make predictions that are in significant tension with existing observations \citep{Kempski:2022,Hopkins:2021}. Two aspects of our simulations represent attractive solutions to the shortcomings of the model in which CRs scatter as a result of turbulence in the ambient medium. First, the magnetic fluctuations in our thermal instability simulation do not exhibit an increase in anisotropy with decreasing scale (\autoref{fig:PS_SF}). This is crucial because anisotropic eddies are very inefficient at scattering CRs \citep{Chandran:2000}. 
Second, plasmoids are regions of extremely high magnetic curvature, so a CR confined to move along a field line that encounters a plasmoid may efficiently scatter if its Larmor radius is comparable to the curvature radius, thus making plasmoids an attractive means to scatter CRs as suggested by \cite{Kempski:2022}.\footnote{Note, that \cite{Kempski:2022} envisioned CR scattering by extremely small, marginally tearing unstable, current sheets that occur in the presence of strong guide fields \citep[e.g.,][]{Boldyrev:2017}, whereas we see much larger current sheets that span a significant fraction of the entire box.}  %% Maybe add an estimate for the size of the plasmoids?? Or say that we have shown that plasmoids are expected to form with sizes of 1e-1 pc and smaller, which is comparable to the larmor radius for 1e5 GeV CRs and smaller. 
Furthermore, the size spectrum of plasmoids may naturally lead to an energy dependent CR diffusion coefficient inline with what is needed to reproduce Milky Way observations \citep{Amato:2018}. We plan to investigate CR scattering in plasmoid dominated MHD turbulence in a future work.

% Cold clump survival
An additional implication of our ISM plasmoid findings is that the lifetime of the smallest structures may be significantly enhanced. The highly anisotropic nature of thermal conduction in the ISM \citep[\eg][]{Spitzer:1962} means that conductive evaporation will be reduced since the interior of a plasmoid will be well insulated due to the fact that, as can be clearly seen in \autoref{fig:Cubes}, it is almost entirely magnetically disconnected from the surrounding medium. We, therefore, expect to find clumps on scales smaller than the standard isotropic Field length \citep{BegelmanMcKee:1990}. This is supported by the findings of thermal instability simulations with anisotropic thermal conduction \citep{Sharma:2010,Choi:2012,Jennings:2021}. Furthermore, the enshrouding magnetic fields will also limit hydrodynamic mixing of the plasmoid with the surroundings \citep{Chandrasekhar:1961}, thereby extending the lifetime and minimum size that should be expected. 

% Turbulence
More generally, our findings demonstrate that we are now getting to the stage where we can move beyond simulations of individual current sheets \citep[\eg][]{Daughton:2011, Huang:2016, Zhang:2021, Beg:2022} and instead study the interaction between turbulence and reconnection \citep[\eg][]{Schekochihin:2020,Nattila:2021,Chernoglazov:2021,Comisso:2022,Dong:2022}. Such studies will help disentangle the intrinsic turbulence \citep[\eg][]{Kowal:2017} that arises from reconnection within a layer and the extrinsically driven turbulence. This will help understand the longevity of plasmoids since, unlike in 2D, in 3D they do not have a clear circular structure, but do exhibit clear coherence in the third dimension. How these plasmoids/flux tubes survive will depend on the interplay of reconnection and turbulence, which will determine how susceptible they are to kinking and dissolving into the background. 

%Future
\subsection{Outlook}
This work represents the beginning of a new line of inquiry into the nature of the magnetized multiphase ISM. There remains much to be done to strengthen and extend our initial findings, and to connect them to observations and other physical processes. We have clearly demonstrated that the plasmoid instability and the associated fast magnetic reconnection is a common occurrence in the interstellar medium and can be induced by thermal instabilities. Furthermore, given the simplicity of our approach our findings are likely to apply to a broad range of systems beyond the ISM, such as the circumgalactic medium, the intracluster medium, and stellar coronae.

\acknowledgements 
We thank Sasha Chernoglazov, Greg Bryan, Eliot Quataert, Chang-Goo Kim, Philipp Kempski, Phil Hopkins, Alisa Galishnikova, Eve Ostriker, and Brent Tan for many useful discussions and suggestions. 
The computations in this work were run at facilities supported by the Scientific Computing Core at the Flatiron Institute, a division of the Simons Foundation.\\
\emph {Software:} \texttt{athena++} \citep{athena++}, \texttt{matplotlib} \citep{matplotlib}, \texttt{scipy} \citep{scipy}, \texttt{numpy} \citep{numpy}, \texttt{ipython} \citep{ipython}, \texttt{CMasher} \citep{CMasher}.

\appendix

\section{Convergence in 3D}\label{sec:converge}

\begin{figure}
    \centering
    \includegraphics[width=\columnwidth]{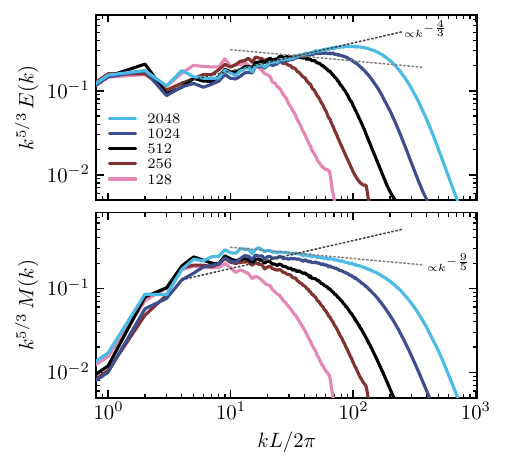}
    \caption{Comparison of the velocity and magnetic field spectra for turbulent thermally instability simulations with resolutions ranging from $L/\Delta x = 2048$ to 128. The approximate power-law fits to the highest resolution simulation are repeated in each panel to ease comparison.}
    \label{fig:convergence PS}
\end{figure}
    
\begin{figure}
    \centering
    \includegraphics[width=\columnwidth]{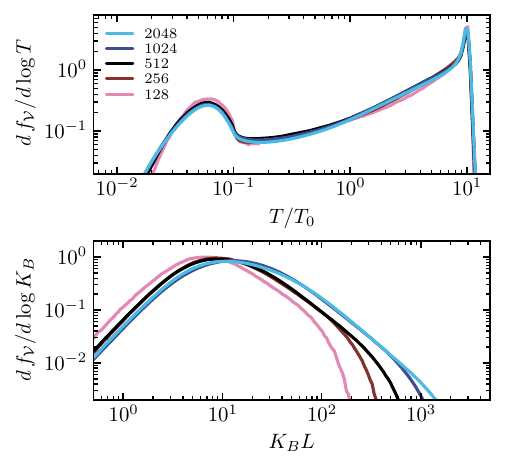}
    \caption{Comparison of the temperature and curvature distributions for turbulent thermally instability simulations with resolutions ranging from $L/\Delta x = 2048$ to 128.}
    \label{fig:convergence Phase}
\end{figure}
    
    \begin{figure}
    \centering
    \includegraphics[width=0.9\columnwidth]{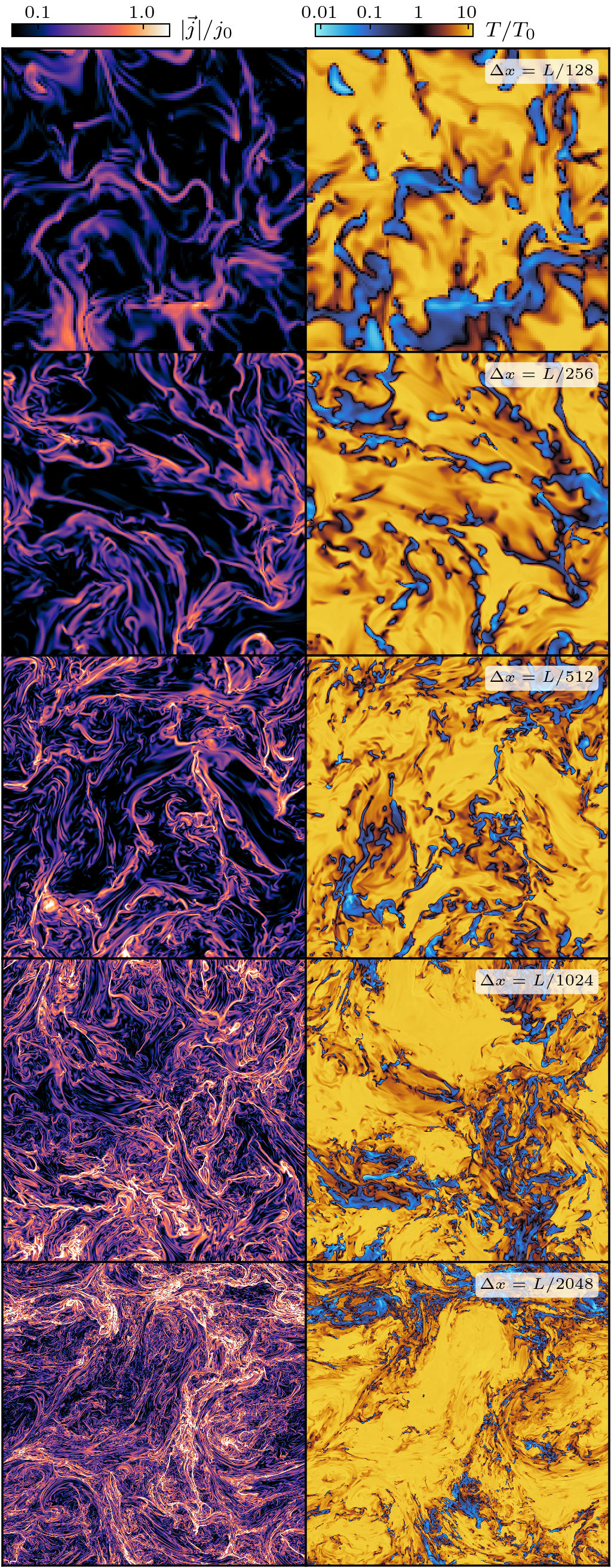}
    \caption{2D slices of current and temperature from simulations with the same parameters as our fiducial simulation but with lower resolution ranging from $\Delta x / L = 1 / 128$ to 1/2048. The current is normalized by $j_0 = B_0 / 512$. A multiphase medium clearly develops, as do large current sheets, however the current sheets in simulations with worse resolution do not show as clear signs of the plasmoid instability because the large numerical dissipation.}
    \label{fig:convergence slice}
\end{figure}

Although on the smallest scales we do not expect our results to be converged since we do not include explicit resolved dissipation processes, on large scales our findings are quite well converged. We demonstrate this by looking at the convergence of the velocity and magnetic field power spectra in \autoref{fig:convergence PS} in which we compare our fiducial turbulent thermally unstable MHD simulation to simulations with up to 16 times lower resolution. The velocity power spectrum, $E(k)$, and magnetic field power spectrum, $M(k)$, on the largest scales are very well converged. At intermediate scales, $4 \lesssim k L / 2 \pi \lesssim 100$, the slope and normalization of $E(k)$ is robust to changes in resolution particularly above $L/\Delta x \geq 512$. $M(k)$ on these same intermediate scales is somewhat less well converged indicating that the increased numerical resistivity at lower resolutions may impact the resolved scales, and that we may be below/just at the critical threshold to form plasmoids in this case. This is consistent with the results from high resolution isothermal resistive MHD simulations \citep{Beresnyak:2014}.%The sum of the velocity and magnetic field spectra is remarkably well converged.

In the top panel of \autoref{fig:convergence Phase} we show that the bulk temperature distribution is extremely well converged with almost no appreciable difference across the full range of resolutions we explored. The bottom panel, on the other hand, shows the curvature distribution, which, not surprisingly, is less well converged with lower resolution simulations tending to have lower curvature on average and less very curved fields. The slope of the curvature distribution, however, is unchanged in simulations with differing resolutions.

Lastly, in \autoref{fig:convergence slice} we give a qualitative picture of the importance of resolution, by showing 2D slices of current, and temperature in simulations with $\Delta x / L = 1 / 128$ to $1/2048$. In all simulations it is clear that large exist but going from low to high resolution the current sheets become far more disturbed and much less laminar. The low resolution simulations show no signs of the plasmoid instability. This highlights the essential need for very high resolution to properly capture the onset of plasmoid formation. We find that our fiducial resolution of $\Delta x / L = 1 / 2048$ is just barely sufficient. The temperature field structure also changes dramatically with resolution with far more small clumps in the higher resolution simulations.

\section{Time independence of statistical Properties}\label{sec:longtime}
\begin{figure}
    \centering
    \includegraphics[width=\columnwidth]{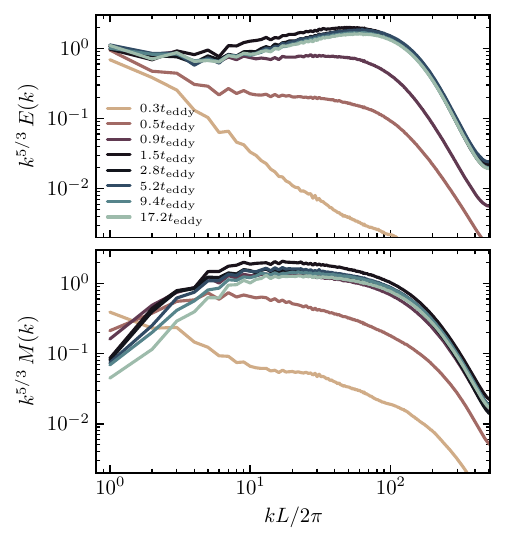}
    \caption{The velocity and magnetic field power spectra at 8 logarithmically spaced times in a turbulent thermally unstable simulation with $\Delta x = L / 1024$. The spectra reach a steady configuration by $\sim 1.5 t_{\rm eddy}$.}
    \label{fig:PowerSpec_vb_time}
    \end{figure}

\begin{figure}
    \centering
    \includegraphics[width=\columnwidth]{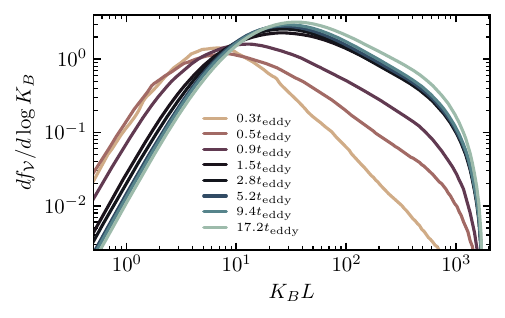}
    \caption{The curvature distribution at 8 logarithmically spaced times in a turbulent thermally unstable simulation with $\Delta x = L / 1024$. The distribution reaches a steady configuration by $\sim 1.5 t_{\rm eddy}$.}
    \label{fig:PDF_K_time}
\end{figure}

For all of our analysis thus far we have focused on measurements occuring at 5 driving-scale eddy turnover times which we approximate as $5 L/v_{\rm rms, warm}$. In this section, we demonstrate that our findings are robust to changes in this arbitrarily chosen measurement time by investigating the time evolution of several key statistical properties in a $\Delta x = L / 1024$ turbulent thermally unstable simulation that we ran for $17.5 L/v_{\rm rms, warm}$, which is 3.5 times longer than our fiducial measurement time. \autoref{fig:PowerSpec_vb_time} and \autoref{fig:PDF_K_time} show the velocity and magnetic power spectra, and the magnetic curvature distributions at 8 logarithmically spaced times, respectively. In both figures it is clear that after  $\sim 1.5 t_{\rm eddy} = L/v_{\rm rms, warm}$ the statistical properties have all reached a time-steady configuration. Therefore, our choice to measure all properties at 5 $t_{\rm eddy}$ is well-justified.


\section{Fully Resolved 2.5D Simulations}\label{sec:2D}

\begin{figure*}
\centering
\includegraphics[width=\textwidth]{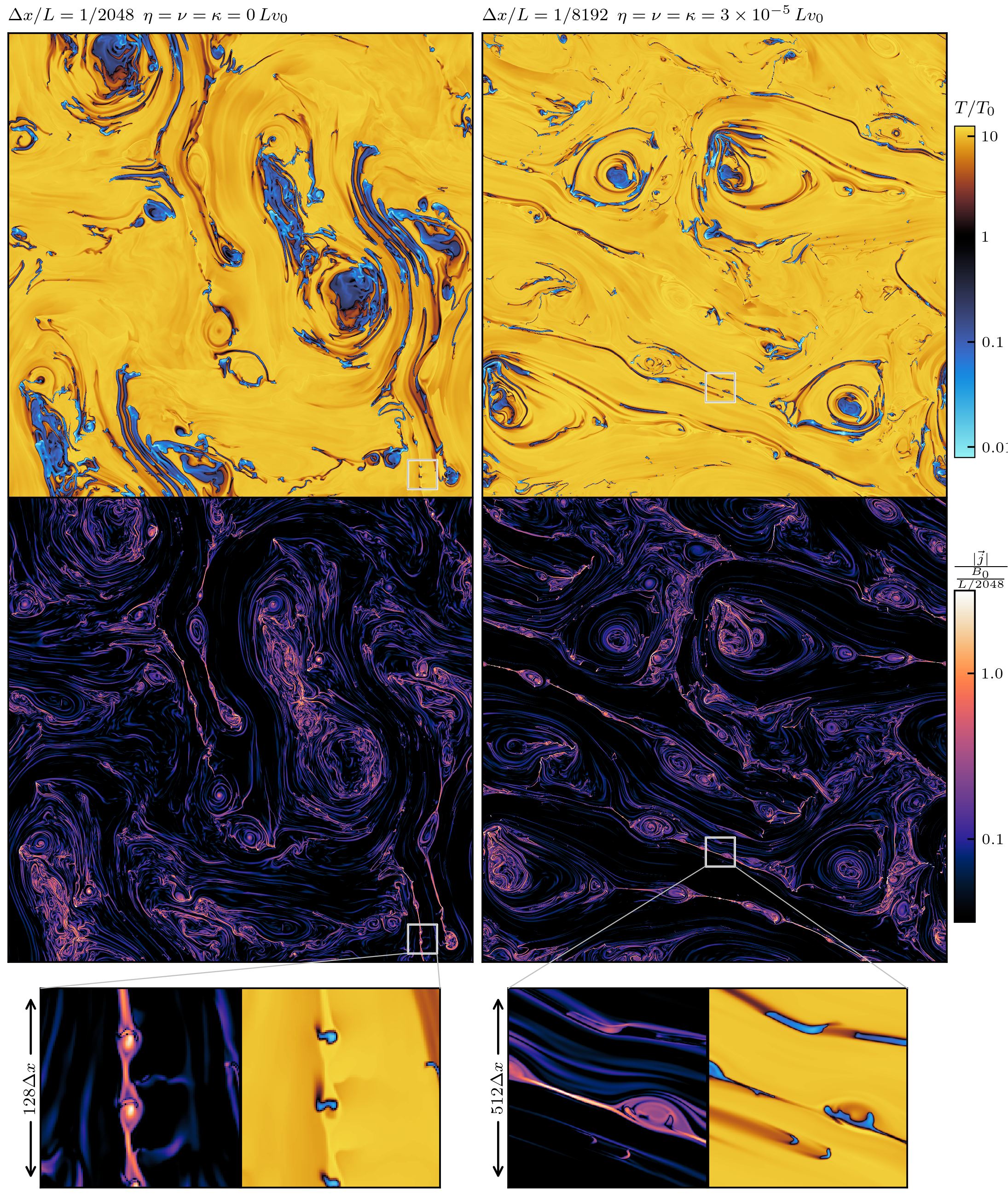}
\caption{Comparison of the temperature (top) and current (bottom) in 2D simulations with either no explicit dissipation processes and a resolution of $\Delta x /L = 1/2048$ (left), or with small but fully resolved dissipation processes and a resolution of $\Delta x /L = 1/8192$ (right). Also show are zoom-ins onto regions 1/16th of the domain centered on tearing unstable current sheets forming cold plasmoids. On both large and small scales there is a marked similarity between these two simulations indicating that numerical dissipation in simulations with high enough resolution can enable plasmoid mediated reconnection in a manner qualitatively consistent with resolved simulations with explicit dissipation. }
\label{fig:2D}
\end{figure*}

Simulations with non-zero explicit resistivity that are high enough resolution to resolve plasmoid unstable current sheets are \textit{extremely} expensive in 3D. As a result, throughout this work we have relied on numerical resistivity. This implicitly assumes that the nature of numerical dissipation is sufficiently similar to explicit resistivity that the plasmoid instability is not changed in important ways. To assess the validity of this simulation we compare a pair of 2.5D\footnote{this means that there are only gradients in two directions but there are three components of the velocity and magnetic field.} simulations with the same conditions as our fiducial simulation. In one of the simulations we adopt the same spatial resolution ($\Delta x /L = 1/2048$) and no explicit dissipation processes, and in the other we adopt 4 times higher resolution ($\Delta x /L = 1/8192$) and explicit resistivity, conductivity, and viscosity ($\eta = \kappa = \nu = 3 \times 10^{-5} L v_0$). This corresponds to Reynolds number of $Re = 3 \times 10^4$, a Prandtl number of $Pr = 1$, and a magnetic Prandtl number of $Pr_M = 1$. At this resolution the dissipation scales are well resolved and sufficiently small to not hinder the onset of the plasmoid instability.

In \autoref{fig:2D} we compare the temperature and current of the two 2.5D simulations. Due to the different nature of turbulence in 2.5D compared to 3D we make this comparison after one eddy turn over time, which is sufficient for the development of fluctuations on all scales and the formation of large current sheets. On a qualitative level it is clear that the thermal and magnetic properties on both large and small scales are very similar. We include zoom-ins on current sheets that are in the process of reconnecting and forming cold plasmoids. These highlight that although the X-points in the lower resolution simulation without explicit dissipation are resolved by only $\sim 4$ cells while in the higher resolution simulation with explicit dissipation they are resolved by $\gtrsim 10$ cells the overall dynamics and the resulting plasmoids are very similar.

\section{The Role of Vorticity}\label{sec:vorticity}
\begin{figure}
    \centering
    \includegraphics[width=\columnwidth]{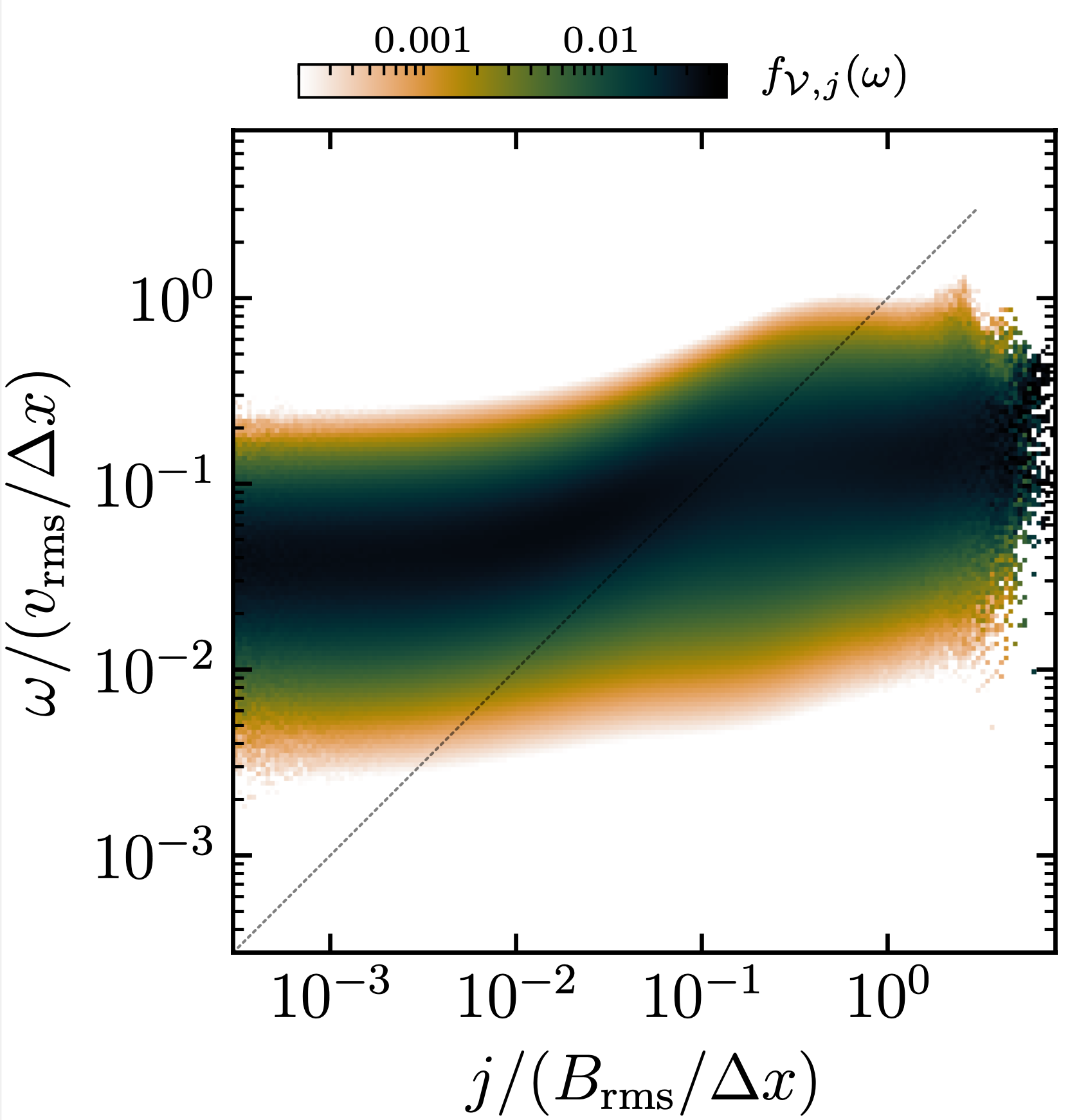}\\
    \includegraphics[width=\columnwidth]{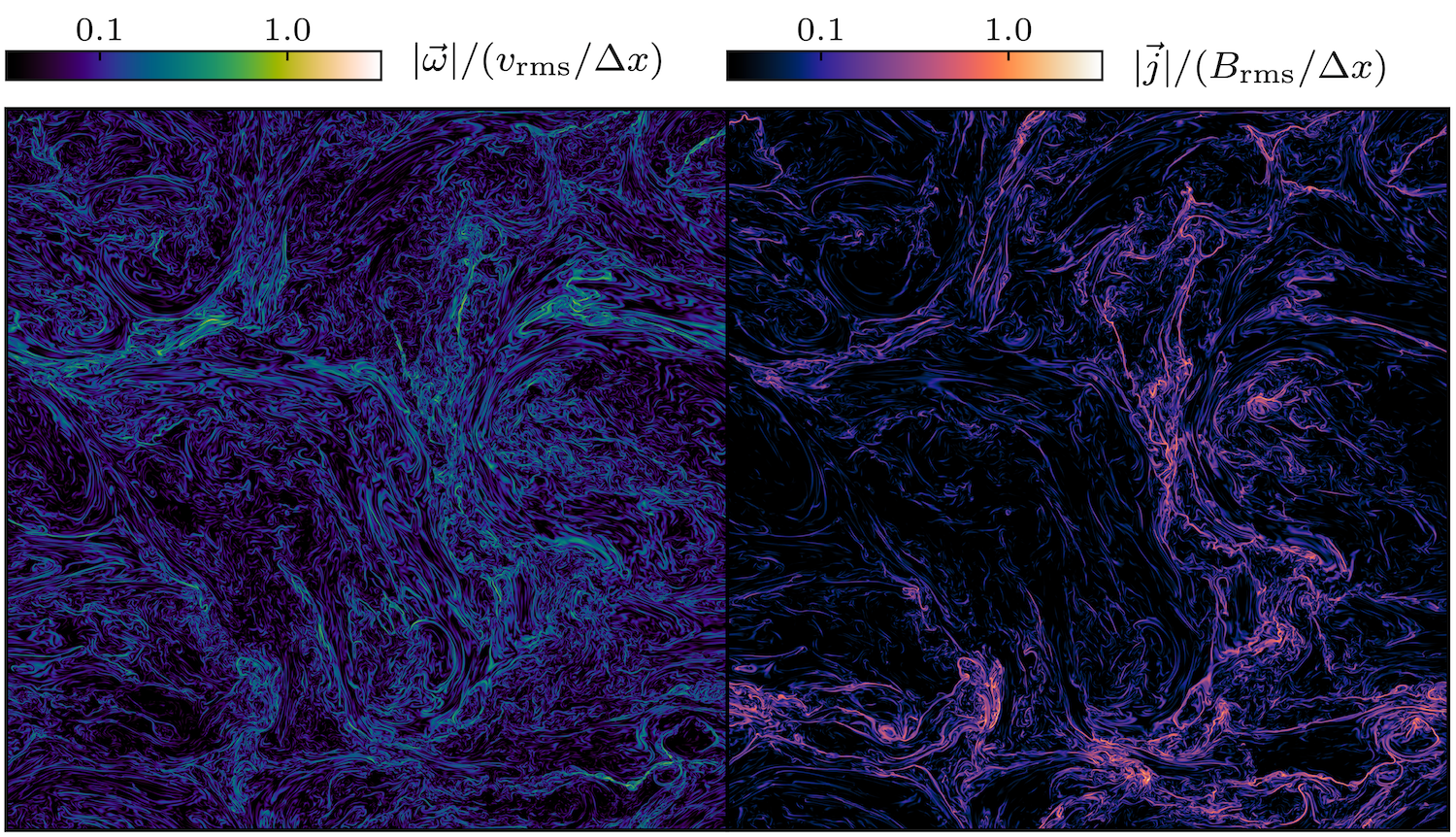}
    \caption{The top panel shows the vorticity distribution in bins of current. The black dashed line shows a one to one relation. This demonstrates that the vorticity varies only slightly over a wide range of current values. The bottom panels show 2D slices of magnitude of the vorticity (left) and current (right). The slices are at the same location as in \autoref{fig:SliceZoom}. There are clear current sheets with small vorticity and some with appreciable vorticity. The plasmoid examples highlighted in \autoref{fig:SliceZoom} exhibit a range of vorticity values ranging from negligible to appreciable.} 
    \label{fig:vorticity}
\end{figure}

To assess the impact of rotational flows on the development of plasmoids we compare the vorticity, $\mathbf{\omega} = \nabla \times \mathbf{v}$, and the current, $\vecj$, in our fiducial simulation in \autoref{fig:vorticity}. It can be clearly seen in these figures that the vorticity varies only slightly over a wide range of current values with only a mild correlation. The maps demonstrate that current sheets, particularly those undergoing plasmoid formation, can have a wide range of vorticities. This suggests that the vorticity does not play a significant role in the development of the plasmoid instability. The nonlinear evolution of the plasmoids may enhance the vorticity in the current sheets.

\section{Cooling and Heating Source Terms}\label{sec:cooling}

\begin{figure}[!b]
\centering
\includegraphics[width=\columnwidth]{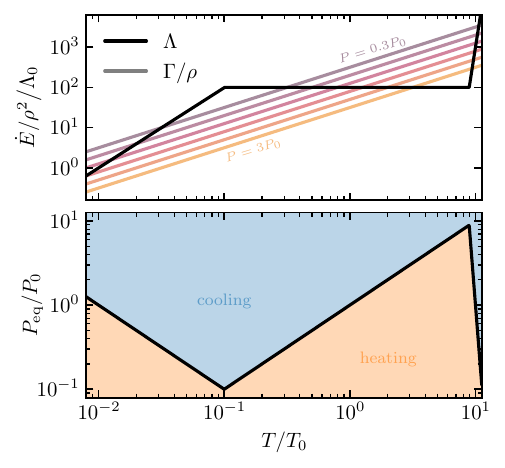}
\caption{(Top) The black line shows the cooling function, and the colored lines show the heating function for a range of pressures. (Bottom) The equilibrium pressure for these cooling and heating functions versus temperature.}
\label{fig:cooling curve}
\end{figure}

The exact functional form of the cooling curve we adopt is given by 
\begin{align}
    \Lambda = \xi \Lambda_0 {\times}
    \begin{cases} 
    (T/T_{\rm lo})^{\alpha_{\rm lo}} & T_{\rm lo} < T \\
    1 & T_{\rm lo} \leq T \leq T_{\rm hi} \\
    (T/T_{\rm hi})^{\alpha_{\rm hi}} & T_{\rm hi} < T
    \end{cases}
\end{align}
where $\xi$ is a dimensionless parameter that controls the strength of cooling, $T_{\rm lo} = T_{\rm cold}^{1 - 1/\alpha_{\rm lo}}$, and $T_{\rm hi} = T_{\rm warm}^{1 - 1/\alpha_{\rm hi}}$. We adopt $\alpha_{\rm lo} = 2$ and $\alpha_{\rm hi} = 20$ to roughly match a Milky Way ISM like cooling curve, and take $T_{\rm cold} = T_0 / 100$, and $T_{\rm warm} = 10 T_0$, which in Milky Way like conditions correspond to roughly 10 K and $10^4$ K, respectively. 

We use a constant heating rate $\Gamma = (P_0 / k_B T_0) \xi \Lambda_0$, which guarantees that $T_{\rm cold}$ and $T_{\rm warm}$ are stable equilibria when $P=P_0$. In our primary simulations we adopt $\xi = 100$, although we also performed several simulations with $\xi = 1$ and 10. This parameter $\xi$ can be thought of as a proxy for the ratio of the eddy turnover timescale to the thermal instability timescale, thus our choice of $\xi \gg 1$ indicates we are in the rapidly thermally unstable regime and can expect there to be a bistable equilibrium phase structure. 

The top panel of \autoref{fig:cooling curve} shows the cooling and heating rates as a function of temperature. The bottom panel shows the equilibrium pressure as a function of temperature.

\bibliographystyle{aasjournal}
\bibliography{references}

\end{document}